\documentclass[aps,preprint,tightenlines,superscriptaddress,showpacs,byrevtex]{revtex4}
\usepackage{graphicx} 
\usepackage{dcolumn}  
\usepackage{amscd}
\usepackage{amsmath}

\graphicspath{{eps/}} 

\begin{document}


\preprint{\vbox{ \hbox{   }
                 \hbox{BELLE-CONF-0743}
}}

\title{ \quad\\[0.5cm]  Measurement of $B \rightarrow D_s K\pi$ branching ratios}

\affiliation{Budker Institute of Nuclear Physics, Novosibirsk}
\affiliation{Chiba University, Chiba}
\affiliation{University of Cincinnati, Cincinnati, Ohio 45221}
\affiliation{Department of Physics, Fu Jen Catholic University, Taipei}
\affiliation{Justus-Liebig-Universit\"at Gie\ss{}en, Gie\ss{}en}
\affiliation{The Graduate University for Advanced Studies, Hayama}
\affiliation{Gyeongsang National University, Chinju}
\affiliation{Hanyang University, Seoul}
\affiliation{University of Hawaii, Honolulu, Hawaii 96822}
\affiliation{High Energy Accelerator Research Organization (KEK), Tsukuba}
\affiliation{Hiroshima Institute of Technology, Hiroshima}
\affiliation{University of Illinois at Urbana-Champaign, Urbana, Illinois 61801}
\affiliation{Institute of High Energy Physics, Chinese Academy of Sciences, Beijing}
\affiliation{Institute of High Energy Physics, Vienna}
\affiliation{Institute of High Energy Physics, Protvino}
\affiliation{Institute for Theoretical and Experimental Physics, Moscow}
\affiliation{J. Stefan Institute, Ljubljana}
\affiliation{Kanagawa University, Yokohama}
\affiliation{Korea University, Seoul}
\affiliation{Kyoto University, Kyoto}
\affiliation{Kyungpook National University, Taegu}
\affiliation{Ecole Polyt\'ecnique F\'ed\'erale Lausanne, EPFL, Lausanne}
\affiliation{University of Ljubljana, Ljubljana}
\affiliation{University of Maribor, Maribor}
\affiliation{University of Melbourne, School of Physics, Victoria 3010}
\affiliation{Nagoya University, Nagoya}
\affiliation{Nara Women's University, Nara}
\affiliation{National Central University, Chung-li}
\affiliation{National United University, Miao Li}
\affiliation{Department of Physics, National Taiwan University, Taipei}
\affiliation{H. Niewodniczanski Institute of Nuclear Physics, Krakow}
\affiliation{Nippon Dental University, Niigata}
\affiliation{Niigata University, Niigata}
\affiliation{University of Nova Gorica, Nova Gorica}
\affiliation{Osaka City University, Osaka}
\affiliation{Osaka University, Osaka}
\affiliation{Panjab University, Chandigarh}
\affiliation{Peking University, Beijing}
\affiliation{University of Pittsburgh, Pittsburgh, Pennsylvania 15260}
\affiliation{Princeton University, Princeton, New Jersey 08544}
\affiliation{RIKEN BNL Research Center, Upton, New York 11973}
\affiliation{Saga University, Saga}
\affiliation{University of Science and Technology of China, Hefei}
\affiliation{Seoul National University, Seoul}
\affiliation{Shinshu University, Nagano}
\affiliation{Sungkyunkwan University, Suwon}
\affiliation{University of Sydney, Sydney, New South Wales}
\affiliation{Tata Institute of Fundamental Research, Mumbai}
\affiliation{Toho University, Funabashi}
\affiliation{Tohoku Gakuin University, Tagajo}
\affiliation{Tohoku University, Sendai}
\affiliation{Department of Physics, University of Tokyo, Tokyo}
\affiliation{Tokyo Institute of Technology, Tokyo}
\affiliation{Tokyo Metropolitan University, Tokyo}
\affiliation{Tokyo University of Agriculture and Technology, Tokyo}
\affiliation{Toyama National College of Maritime Technology, Toyama}
\affiliation{Virginia Polytechnic Institute and State University, Blacksburg, Virginia 24061}
\affiliation{Yonsei University, Seoul}
  \author{K.~Abe}\affiliation{High Energy Accelerator Research Organization (KEK), Tsukuba} 
  \author{I.~Adachi}\affiliation{High Energy Accelerator Research Organization (KEK), Tsukuba} 
  \author{H.~Aihara}\affiliation{Department of Physics, University of Tokyo, Tokyo} 
  \author{K.~Arinstein}\affiliation{Budker Institute of Nuclear Physics, Novosibirsk} 
  \author{T.~Aso}\affiliation{Toyama National College of Maritime Technology, Toyama} 
  \author{V.~Aulchenko}\affiliation{Budker Institute of Nuclear Physics, Novosibirsk} 
  \author{T.~Aushev}\affiliation{Ecole Polyt\'ecnique F\'ed\'erale Lausanne, EPFL, Lausanne}
                    \affiliation{Institute for Theoretical and Experimental Physics, Moscow} 
  \author{T.~Aziz}\affiliation{Tata Institute of Fundamental Research, Mumbai} 
  \author{S.~Bahinipati}\affiliation{University of Cincinnati, Cincinnati, Ohio 45221} 
  \author{A.~M.~Bakich}\affiliation{University of Sydney, Sydney, New South Wales} 
% ...
  \collaboration{The Belle Collaboration}
\noaffiliation

\begin{abstract}
We report a measurement of the exclusive $B^+$ meson decay to the
final state $D_s^- K^+\pi^+$ using $520 \times 10^{6} B\bar{B}$ pairs
collected near the $\Upsilon(4S)$ resonance, with the Belle detector
at the KEKB asymmetric-energy $e^+e^-$ collider. Using the $D_S^- \to
\phi \pi^-$ decay mode to reconstruct $D_s^-$ mesons, we obtain the
branching fraction ${\cal B}(B^+\to D_s^-K^+\pi^+)=
(1.77^{+0.12}_{-0.12} {\mathrm (stat)} \pm 0.16 {\mathrm (syst)} \pm
0.23 {\cal(B) })\times 10^{-4}$. We also present preliminary results
of a study of the two-body $D_sK$, $D_s\pi$ and $K\pi$ subsystems
observed in this final state.
\end{abstract}

\pacs{13.25.Hw, 14.40.Nd}

\maketitle


{\renewcommand{\thefootnote}{\fnsymbol{footnote}}}
\setcounter{footnote}{0}


\section{Introduction}

\begin{figure}[htb]
\centering
\includegraphics[height=5.0cm]{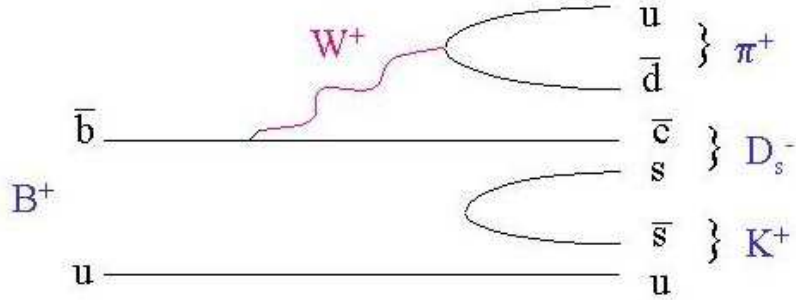}
\caption{\it Diagram for the decay $B^+\to D_s^- K^+\pi^+$.}
\label{RYS1}
\end{figure}
\begin{figure}[htb]
\centering
\includegraphics[height=5.0cm]{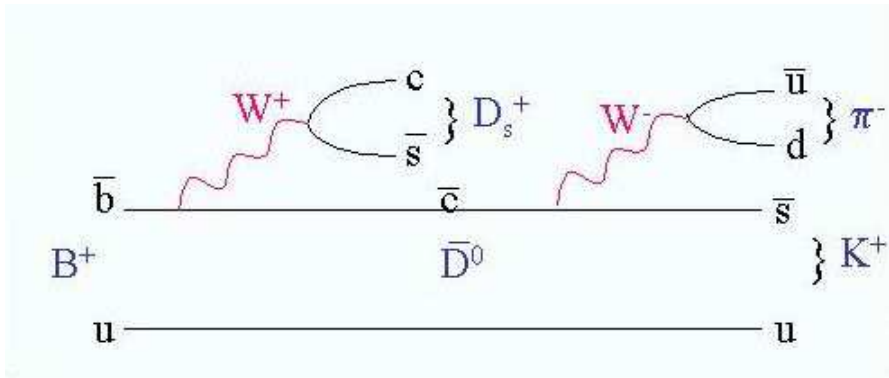}
\caption{\it Diagram for the decay $B^+\to D_s^+ \bar{D^0}, \bar{D^0}\to K^+\pi^-$.}
\label{RYS2}
\end{figure}

We search for the exclusive decays of charged $B$ mesons into $D_s K
\pi$ final states. These modes offer rich possibilities for studies of
different two-body subsystems, such as $D_s K$. For the first decay
mode studied, $B^+\to D_s^- K^+\pi^+$~\cite{FOOT}, the dominant
process is described by the Feynman diagram shown in
Fig.~\ref{RYS1}. This process is mediated by the $b\to c$ quark
transition and includes the production of an $s\bar{s}$ pair via a
radiative gluon. The $D_s K \pi$ final state can also be a result of
the $b\to c \to s$ decay chain, where the $D_s$ meson and the charged
pions are produced in two $W$ boson decays. The Feynman diagram in
Fig.~\ref{RYS2} describes the dominant process responsible for the
second decay mode, $B^+\to D_s^+ \bar{D^0}$ with $\bar{D^0}\to
K^+\pi^-$. Note that the two different processes lead to a similar
three-body final state, but with opposite charges for the $D_s$ and
$\pi$ mesons. We measure the branching fraction of the first decay
mode and study two-body subsystems, while the second decay mode is a
control channel to check the reliability of the method.

In the following the Belle detector and the data sample are briefly
described. Next we discuss the reconstruction of the $D_s K \pi$
final states. Finally the determination of the branching fractions
together with a preliminary study of the $D_sK\pi$ subsystems
including $D_sK $, $D_s\pi$ and $K\pi$ are presented.

\section{Detector and data Sample}


The results are based on a data sample that contains $(520.2 \pm 6.8)
\times 10^6 B\bar{B}$ pairs, corresponding to an integrated
luminosity of 479~fb$^{-1}$, collected with the Belle detector at the
KEKB asymmetric-energy $e^+e^-$ (3.5 on 8~GeV) collider
\cite{KEKB}. KEKB operates at the $\Upsilon(4S)$ resonance
($\sqrt{s}=10.58$~GeV) with a peak luminosity that exceeds $1.7 \times
10^{34}~{\rm cm}^{-2}{\rm s}^{-1}$. The $\Upsilon(4S)$ resonance is
produced with a Lorentz boost of $\beta\gamma=0.425$ nearly along the
electron beamline ($z$-axis).  The production rates of $B^+B^-$ and
$B^0\overline{B}{}^0$ pairs are assumed to be equal.

The Belle detector is a large-solid-angle magnetic spectrometer that
consists of silicon vertex detector (SVD), a 50-layer central drift
chamber (CDC), an array of aerogel threshold \v{C}erenkov counters
(ACC), a barrel-like arrangement of time-of-flight scintillation
counters (TOF), and an electromagnetic calorimeter composed of CsI(Tl)
crystals (ECL) located inside a super-conducting solenoid coil that
provides a 1.5~T magnetic field. An iron flux-return located outside
of the coil is instrumented to detect $K_L^0$ mesons and to identify
muons (KLM). The detector is described in detail
elsewhere~\cite{BELLE}. Two inner detector configurations were
used. A 2.0 cm beampipe and a 3-layer silicon vertex detector was used
for the first sample of $152 \times 10^6 B\bar{B}$ pairs, while a 1.5
cm beampipe, a 4-layer silicon detector and a small-cell inner drift
chamber were used to record the remaining $368 \times 10^6 B\bar{B}$
pairs \cite{SVD}.


\section{Reconstruction of $B^+ \rightarrow D_s^- K^+\pi^+$ and $B^+\to D_s^+ \bar{D^0}$}


The reconstruction of $B^+ \to D_s^- K^+\pi^+$ and $B^+\to D_s^+
\bar{D^0}$ decays consists of the following steps: selection of
charged tracks, discrimination between $B\bar{B}$ and continuum
events, particle identification and reconstruction of all intermediate
decays in $B \rightarrow D_s K\pi$ processes. Each of these steps is
briefly described below.

\subsection{Track selection}

Charged tracks with $dr <$ 0.5 cm and $dz <$ 5 cm are selected,
where $dr$ and $dz$ are impact parameters measured in the $r$-$\phi$
(transverse) plane and $z$ direction, respectively. Charged tracks are
also required to have transverse momenta greater than $100~{\rm
MeV}/c$.

\subsection{Suppression of continuum events}

We exploit the event topology to discriminate between spherical
$B\bar{B}$ events and the dominant background from jet-like continuum
events, $e^{+}e^{-} \rightarrow q\bar{q}$ ($q$ = $u$, $d$, $s$,
$c$). We combine event shape variables using all particles in an event
to calculate Fox-Wolfram moments \cite{FOX} and define the $R_2$, ratio of 
second and zeroth moments, as
\begin{equation*}
  R_2 = 
  \frac{\sum_{i,j} |\vec{p}_i| |\vec{p}_j| P_2 (\cos \theta_{ij}) }
       {\sum_{i,j} |\vec{p}_i| |\vec{p}_j|} \, ,
\end{equation*}
where $\vec{p}$ indicates the particle momentum, $P_2$ is the Legendre
polynomial of second order, and $i$, $j$ enumerate all particles in the
event. We require $R_2$ to be less than 0.4.

\subsection{Particle identification}

Hadron identification is based on information provided by the CDC, ACC
and TOF. For kaons we require ${\cal L}_{K/\pi}>0.6$ and ${\cal
L}_{p/K} < 0.95$ (veto), where ${\cal L}_{x/y} = \frac{{\cal
L}_x}{{\cal L}_x + {\cal L}_y}$ ($x$, $y$ = $\pi$, $K$, $p$) denotes
the corresponding likelihood ratio. Pions are selected as non-kaons,
satisfying veto conditions for $K$ and $p$: ${\cal L}_{p/K} < 0.95$
and ${\cal L}_{K/\pi} < 0.95$. In addition, we reject tracks that are
consistent with an electron hypothesis. A selection imposed on this
ratio results in a typical kaon (pion) identification efficiency
ranging from 92\% to 97\% (94\% to 98\%) for various decay modes,
while 2\% to 15\% (4\% to 8\%) of kaon (pion) candidates are
misidentified pions (kaons).

\subsection{Reconstruction of exclusive $D_s^+$ decays}

The $D_s^+$ candidates are reconstructed in two final states:
$\phi(\to K^+K^-)\pi^+ $ and $\bar{K^*}(892)^0(\to K^-\pi^+) K^+ $.
We reconstruct $\phi$ mesons in the $ K^+K^-$ final state.  The $\phi$
mass is estimated to be ($1019.70 \pm 0.153$) MeV/$c^2$, in agreement
with the world average value $m_{WA}(\phi) = (1019.460\pm 0.019$)
MeV/$c^2$ ~\cite{PDG}.  We accept $K^+K^-$ pairs satisfying the
requirement $\left| m(K^+K^-) - m_{WA}(\phi)\right| < 10$ MeV/$c^2$.
Similarly, the $K^{+}\pi^{-}$ mass spectrum exhibits a $K^{*}(892)^0$
signal at a mass of ($892.19 \pm 1.80$) MeV/$c^2$
($m_{WA}(K^{*0})=(896.00\pm 0.25$)~MeV/$c^2$ ~\cite{PDG}); $K^{*0}$
candidates are thus selected with the requirement: $\left| m(K^+\pi^-)
- m_{WA}(K^{*0})\right| < 100$ MeV/$c^2$.

\begin{figure}[hbt]
\setlength{\unitlength}{1mm}
\begin{center}
\begin{picture}(120,80)
\put(75,0){\large $m(\phi\pi^-)$[GeV/$c^2$]}
\put(6,77){\large\bf $\frac{dN}{d\;m(\phi\pi^-)\;\cdot\;(0.002\;\mathrm{GeV}/c^2)}$}
\includegraphics[height=8cm,width=12cm]{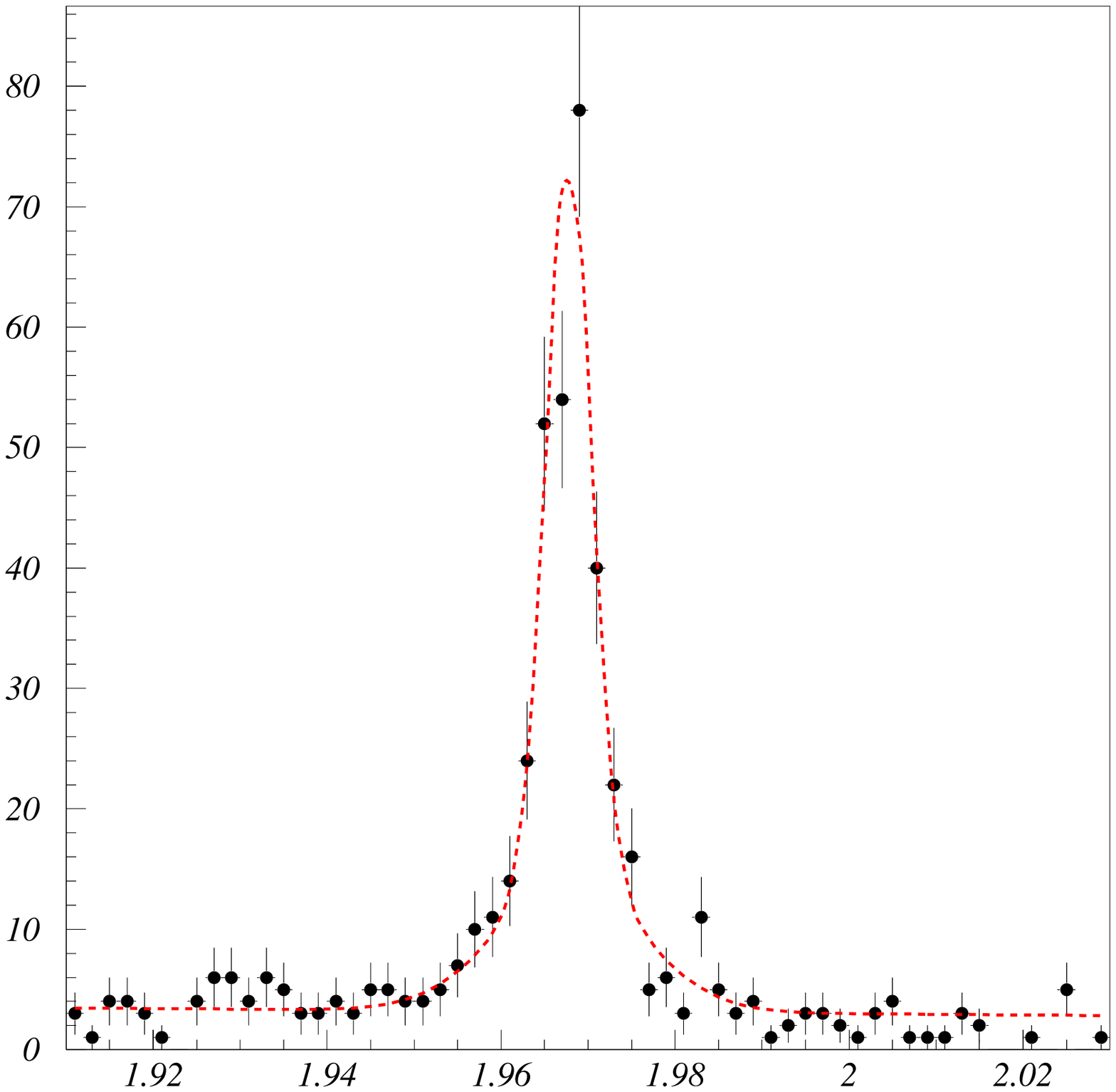}
\end{picture}
\end{center}
\caption{\it Invariant mass distribution for $\phi\pi^-$ pairs (points)
  together with the curve representing the fit result in the range of
  the $D_s^-$ meson mass (red dashed line) for $B^+ \rightarrow D_s^{-}
  K^+\pi^+$ decay. The signal is parameterized by two Gaussians, while 
  the background is described by a second-order polynomial. The
  spectrum corresponds to the ($\Delta E, M_\mathrm{bc}$) signal box.}
\label{FIGDS1}
\setlength{\unitlength}{1mm}
\begin{center}
\begin{picture}(120,80)
\put(57,0){\large $m(K^*(892)^0 K^-)$[GeV/$c^2$]}
\put(6,77){\large\bf $\frac{dN}{d\;m(K^*(892)^0 K^-)\;\cdot\;(0.002\;\mathrm{GeV}/c^2)}$}
\includegraphics[height=8cm,width=12cm]{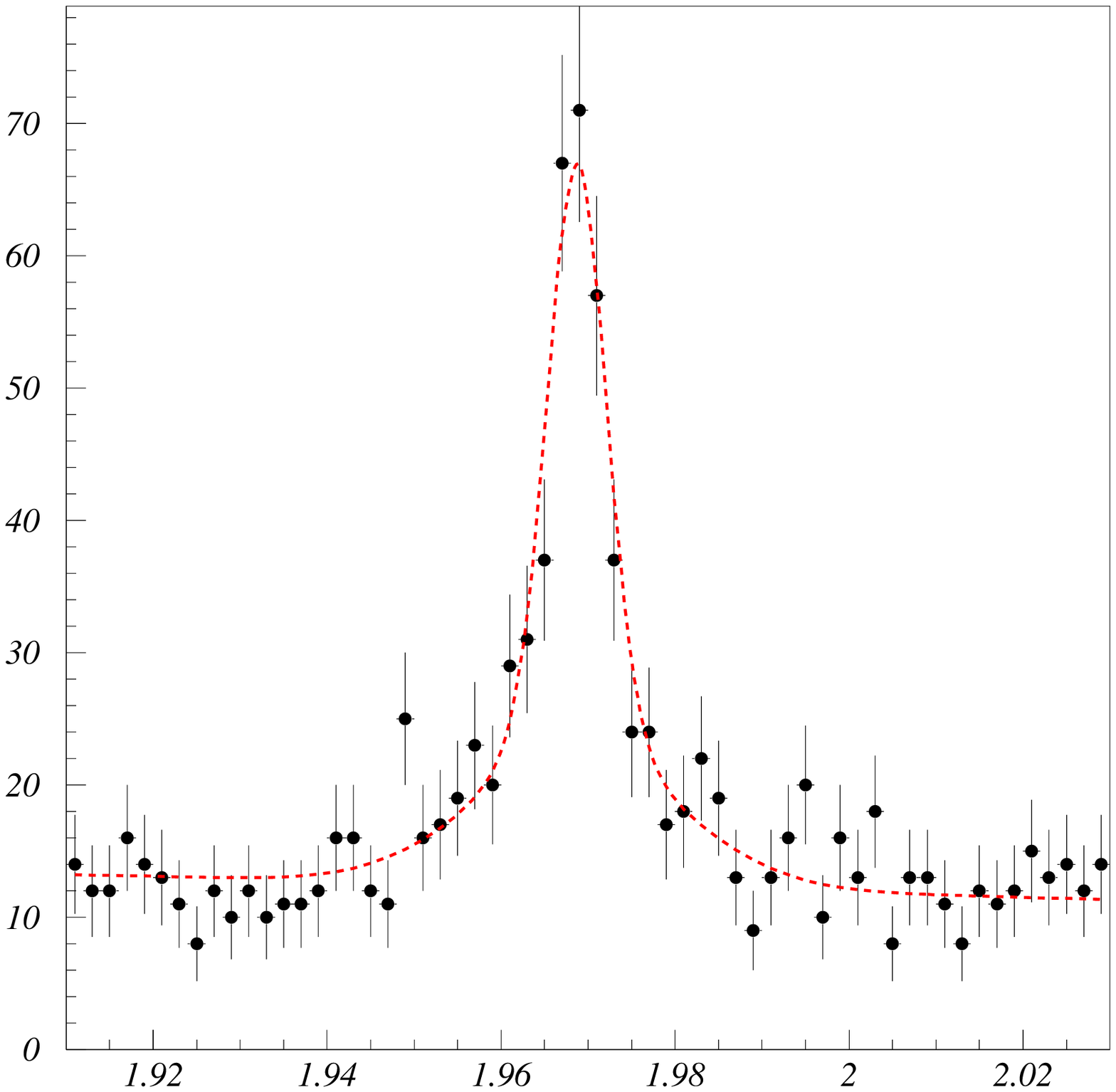}
\end{picture}
\end{center}
\caption{\it Invariant mass distribution for $K^*(892)^0 K^-$ pairs
  (points) together with the curve representing the fit result 
  in the range of the $D_s^-$ meson mass (red dashed line) for 
  $B^+ \rightarrow D_s^{-} K^+\pi^+$ decay. The signal is parameterized
  by two Gaussians, while the background is described by a second-order
  polynomial. The spectrum corresponds to the ($\Delta E, M_\mathrm{bc}$) signal box.}
\label{FIGDS2}
\end{figure}

A clear $D_s$ meson signal is observed in both channels considered
(Figs. \ref{FIGDS1}--\ref{FIGDS2}). The fitting procedure -- described
in the next section -- is used to obtain the mean values of the $D_s$ mass
in each channel.  This mass, averaged over two decay channels studied,
is estimated to be ($1968.1 \pm 0.20 {\mathrm (stat)}$) MeV/$c^2$,
which is consistent with the world average value $m_{WA}(D_s)=(1968.2
\pm 0.5)$ MeV/$c^2$ ~\cite{PDG}. The invariant mass of $D_s$
candidates is required to satisfy the criterion $\left| m(D_s) -
m_{WA}(D_s)\right| < 15$ MeV/$c^2$, corresponding to a window of
approximately three standard deviations about the $D_s$ mass.

\subsection{Reconstruction of $B$ mesons}

$B$ mesons are reconstructed combining $D_s$ candidates with
identified kaons and pions. In exclusive reconstruction of $B$ mesons
two kinematic variables are used: the energy difference, $\Delta E$,
and the beam-energy-constrained mass, $M_\mathrm{bc}$. These are defined as:
\begin{equation}
  \Delta E = E_B - E_\mathrm{beam}
\end{equation}
and
\begin{equation}
  M_\mathrm{bc} = \sqrt{E_\mathrm{beam}^2 - p_B^2} \, ,
\end{equation}
where $E_B$ and $P_B$ are the reconstructed energy and momentum of the
$B$ candidate, and $E_\mathrm{beam}$ is the run-dependent beam energy,
all expressed in the centre-of-mass (CM) frame. We use the dedicated
Monte Carlo (MC) sample of $B \rightarrow D_s K\pi$ decays to define
signal regions in $\Delta E$ and $M_\mathrm{bc}$ as $|\Delta E| <
0.03$ GeV and $M_\mathrm{bc} > 5.27$ GeV/$c^2$. For all $B$
candidates, a loose requirement on the goodness of the $D_s K\pi$
vertex fit -- with the $D_s$ mass constrained to the world average
value -- is also applied ($\chi^2_B < 50$).

From MC simulation we determine that the background contribution from
$B^+\to ((c\bar{c})\to K^+K^-\pi^+\pi^-) K^+$ decays, where
$(c\bar{c})$ are charmonium states such as the $J/\psi$ or $\eta_c$, is removed by
discarding all events that satisfy the requirement: 2.88 GeV/$c^2 <
M(K^+K^-\pi^+\pi^-) <$ 3.18 GeV/$c^2$.

\begin{figure}[p]
\setlength{\unitlength}{1mm}
\begin{center}
\begin{picture}(120,80)
\put(67,0){\large $m(K^{+}\pi^{-})$ [GeV/$c^2$]}
\put(6,78){\large\bf $\frac{dN}{d\;m(K^+\pi^-)\;\cdot\;(0.002\;\mathrm{GeV}/c^2)}$}
\includegraphics[height=8cm,width=12cm]{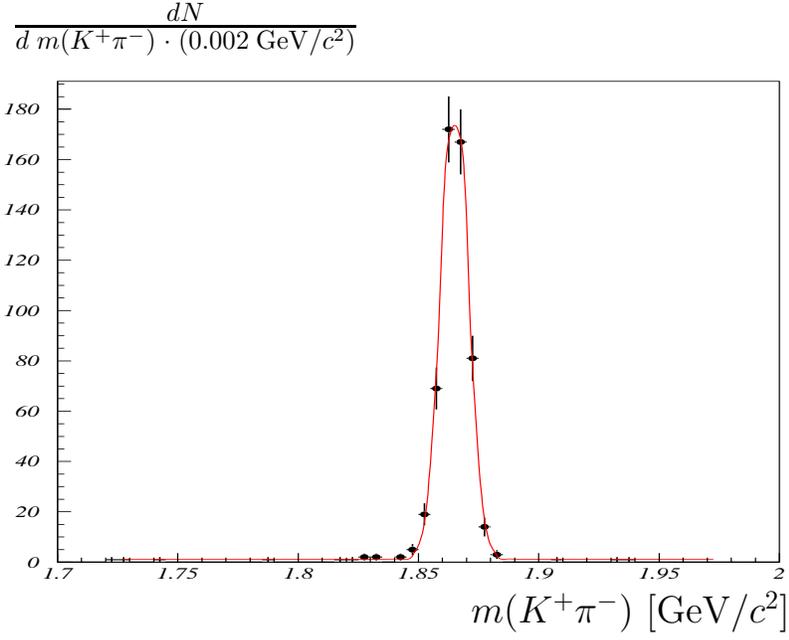}
\end{picture}
\end{center}
\caption{\it Invariant mass distribution for $K^{+}\pi^{-}$ pairs
  (points) together with the curve representing the fit result in the
  range of the $\bar{D}^0$ meson for the $B^+\rightarrow
  D_s^+\bar{D}^0$ decay. The signal is parameterized by a single
  Gaussian, while the background is described by a first-order
  polynomial. The spectrum corresponds to the ($\Delta E, M_\mathrm{bc}$) signal box.}
\label{FIGD0}
\end{figure}

For $B^+ \rightarrow D_s^+ \bar{D^0}$ candidate events a requirement
on $K^+\pi^-$ invariant mass is imposed: $|m(K^+\pi^-) - m_{WA}(D^0)|
< 0.015$ GeV/$c^2$~\cite{PDG}. This condition is determined by using
$D_s^+K^+\pi^-$ combinations for events from the ($\Delta E, M_\mathrm{bc}$)
signal region and plotting the invariant mass of $K^+\pi^-$ pairs as
shown in Fig. \ref{FIGD0}.

After selection requirements, 8.8\%(10.6\%) of events with
reconstructed $D_s \to \phi\pi$ ($D_s \to K^{*0}K$) decays have more
than one $B$ candidate. In these cases we select the $B$ candidate
having the smallest value of $\chi_B^2$. If there are two or more
combinations with the same value of $\chi^2_B$, the one containing a
kaon -- originating directly from the $B$ decay -- with the highest
likelihood ratio ${\cal L}_{K/\pi}$ is selected.

\begin{figure}[t]
\setlength{\unitlength}{1mm}
\begin{center}
\begin{picture}(130,90)
\put(88,0){\large $\Delta E$[GeV]}
\put(10,77){\large\bf $\frac{dN}{d(\Delta E)\;\cdot\;(0.008\;\mathrm{GeV})}$}
\includegraphics[height=8cm,width=12cm]{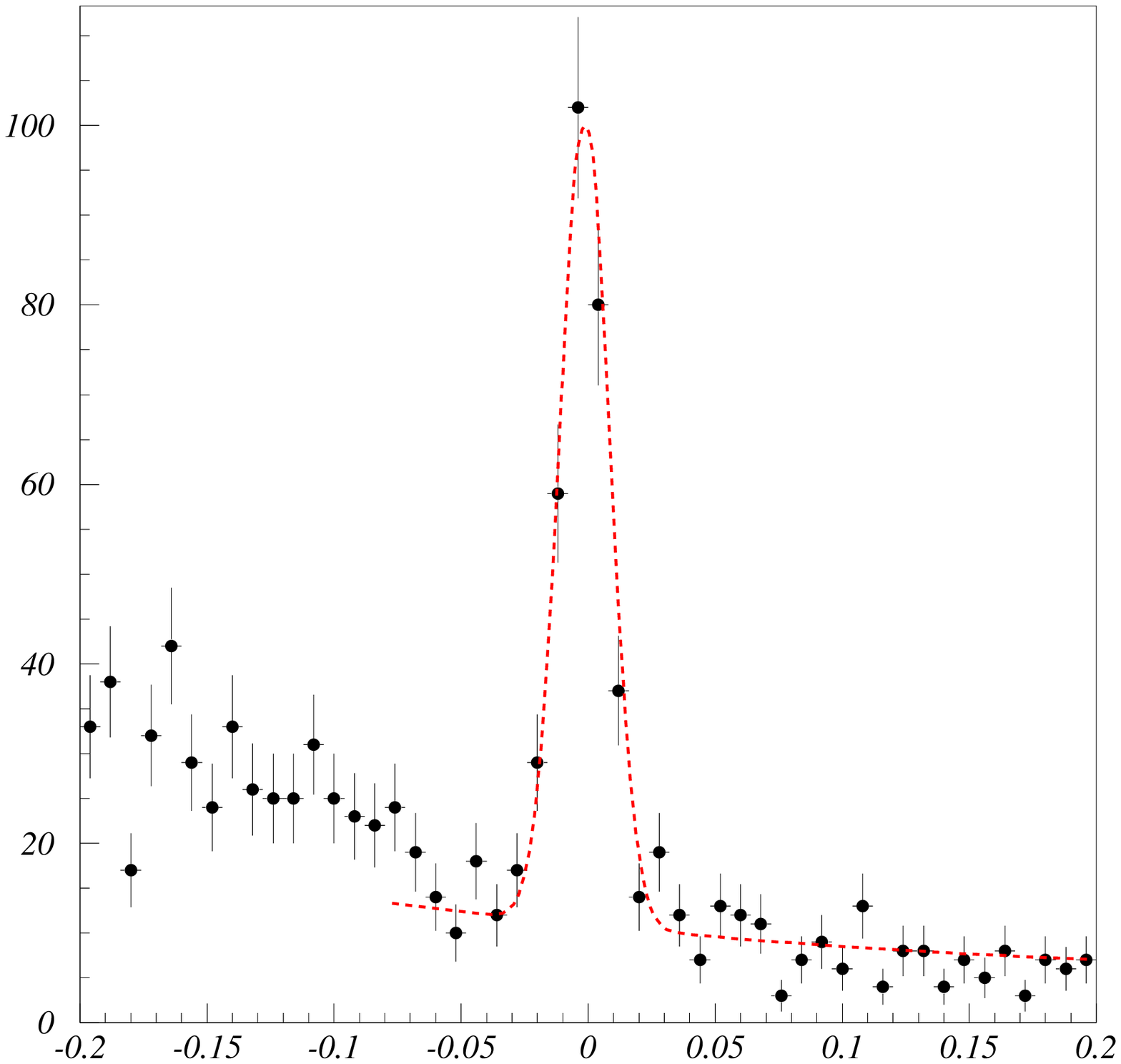}
\end{picture}
\end{center}
\setlength{\unitlength}{1mm}
\begin{center}
\begin{picture}(130,90)
\put(80,0){\large $M_\mathrm{bc}$[GeV/$c^2$]}
\put(10,77){\large\bf $\frac{dN}{dM_\mathrm{bc}\;\cdot\;(0.002\;\mathrm{GeV}/c^2)}$}
\includegraphics[height=8cm,width=12cm]{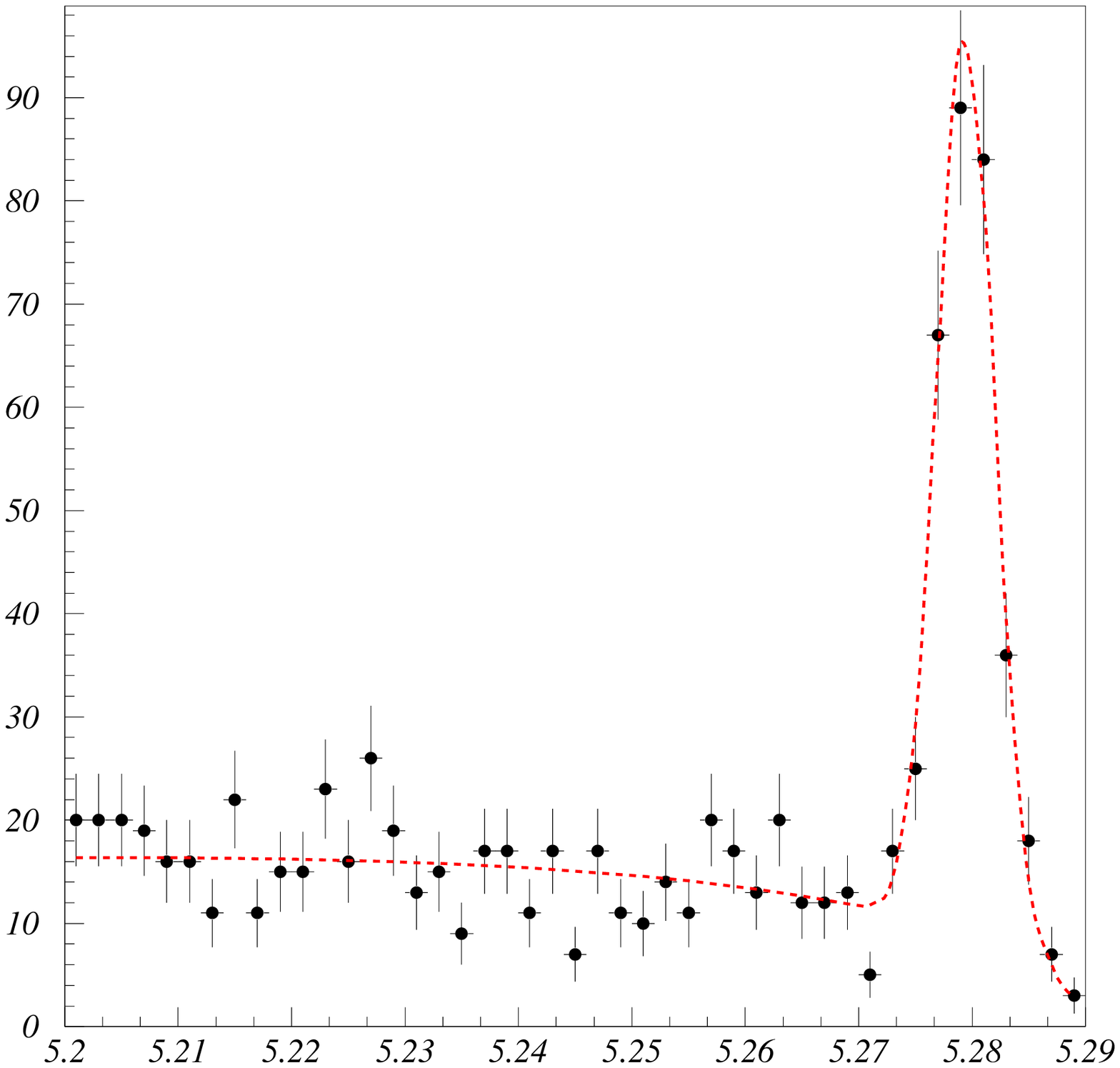}
\end{picture}
\end{center}
\caption{\it Distributions of the $\Delta E$ (top) and $M_\mathrm{bc}$ variables (bottom) for the decay 
$B^{+} \rightarrow D_s^{-} K^{+} \pi^{+}$, $D_s^-\to \phi
\pi^-$. Points correspond to data and the dashed (red) line
represents results of the fit described in the text. }
\label{FIG_DSKAPI_1}
\end{figure}

\begin{figure}[t]
\setlength{\unitlength}{1mm}
\begin{center}
\begin{picture}(130,90)
\put(88,0){\large $\Delta E$[GeV]}
\put(10,77){\large\bf $\frac{dN}{d(\Delta E)\;\cdot\;(0.008\;\mathrm{GeV})}$}
\includegraphics[height=8cm,width=12cm]{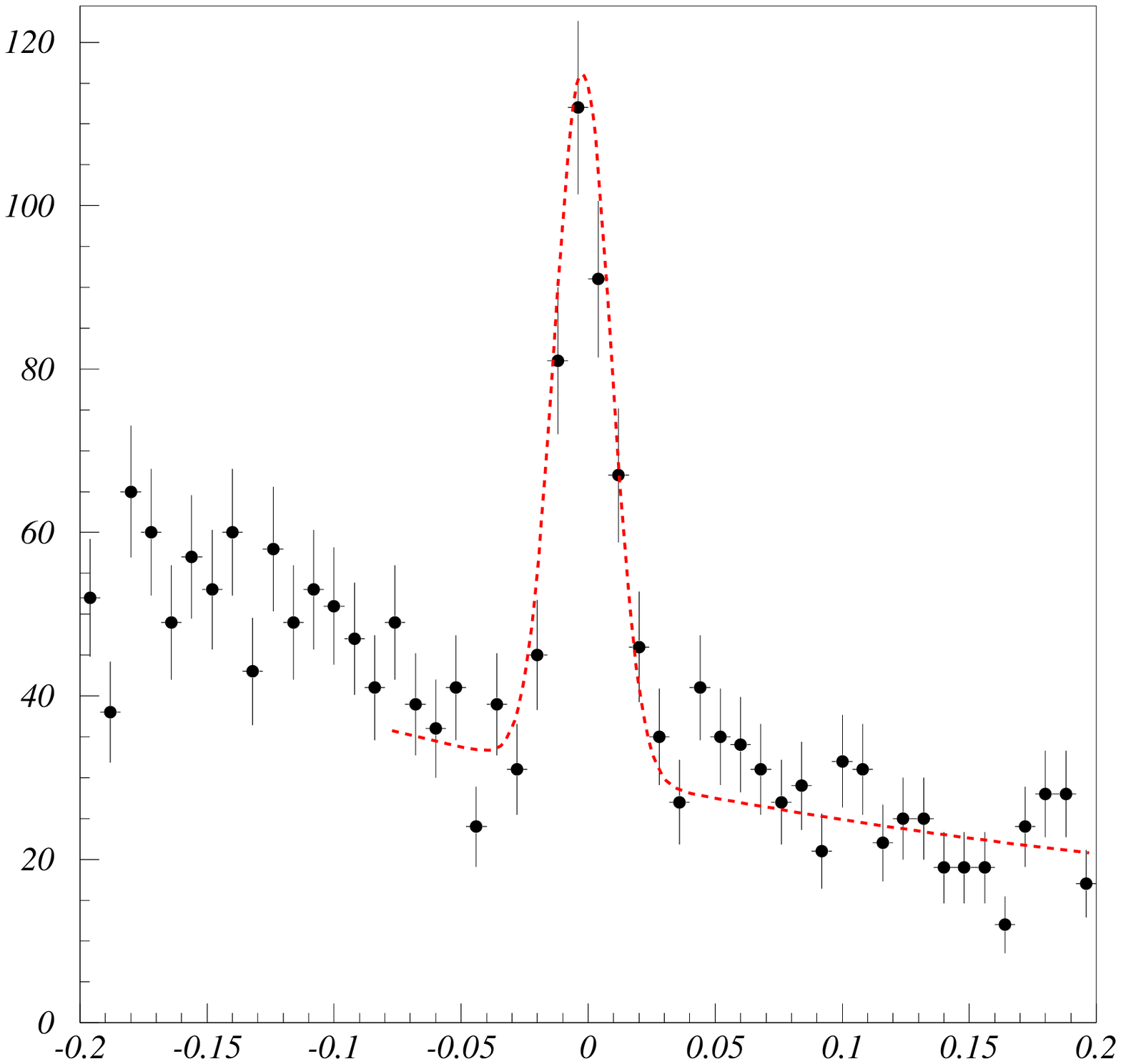}
\end{picture}
\end{center}
\setlength{\unitlength}{1mm}
\begin{center}
\begin{picture}(130,90)
\put(80,0){\large $M_\mathrm{bc}$[GeV/$c^2$]}
\put(10,77){\large\bf $\frac{dN}{dM_\mathrm{bc}\;\cdot\;(0.002\;\mathrm{GeV}/c^2)}$}
\includegraphics[height=8cm,width=12cm]{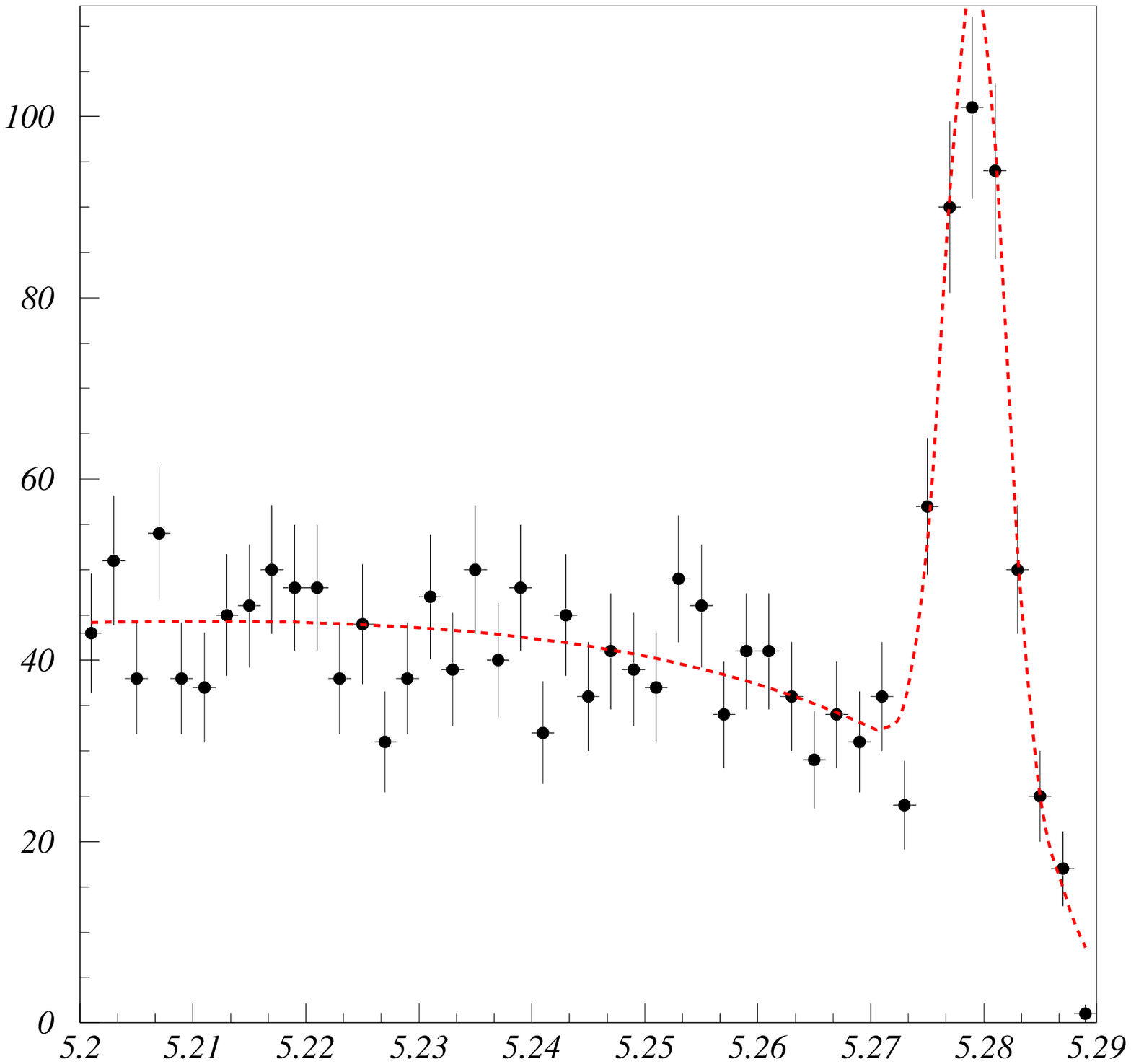}
\end{picture}
\end{center}
\caption{\it Distributions of the $\Delta E$ (top) and $M_\mathrm{bc}$ variables (bottom) for the decay 
$B^{+} \rightarrow D_s^{-} K^{+} \pi^{+}$, $D_s^-\to K^*(892)^0
K^-$. Points correspond to data and the dashed (red) line
represents results of the fit described in the text. }
\label{FIG_DSKAPI_2}
\end{figure}

\begin{figure}[t]
\setlength{\unitlength}{1mm}
\begin{center}
\begin{picture}(130,90)
\put(88,0){\large $\Delta E$[GeV]}
\put(10,77){\large\bf $\frac{dN}{d(\Delta E)\;\cdot\;(0.008\;\mathrm{GeV})}$}
\includegraphics[height=8cm,width=12cm]{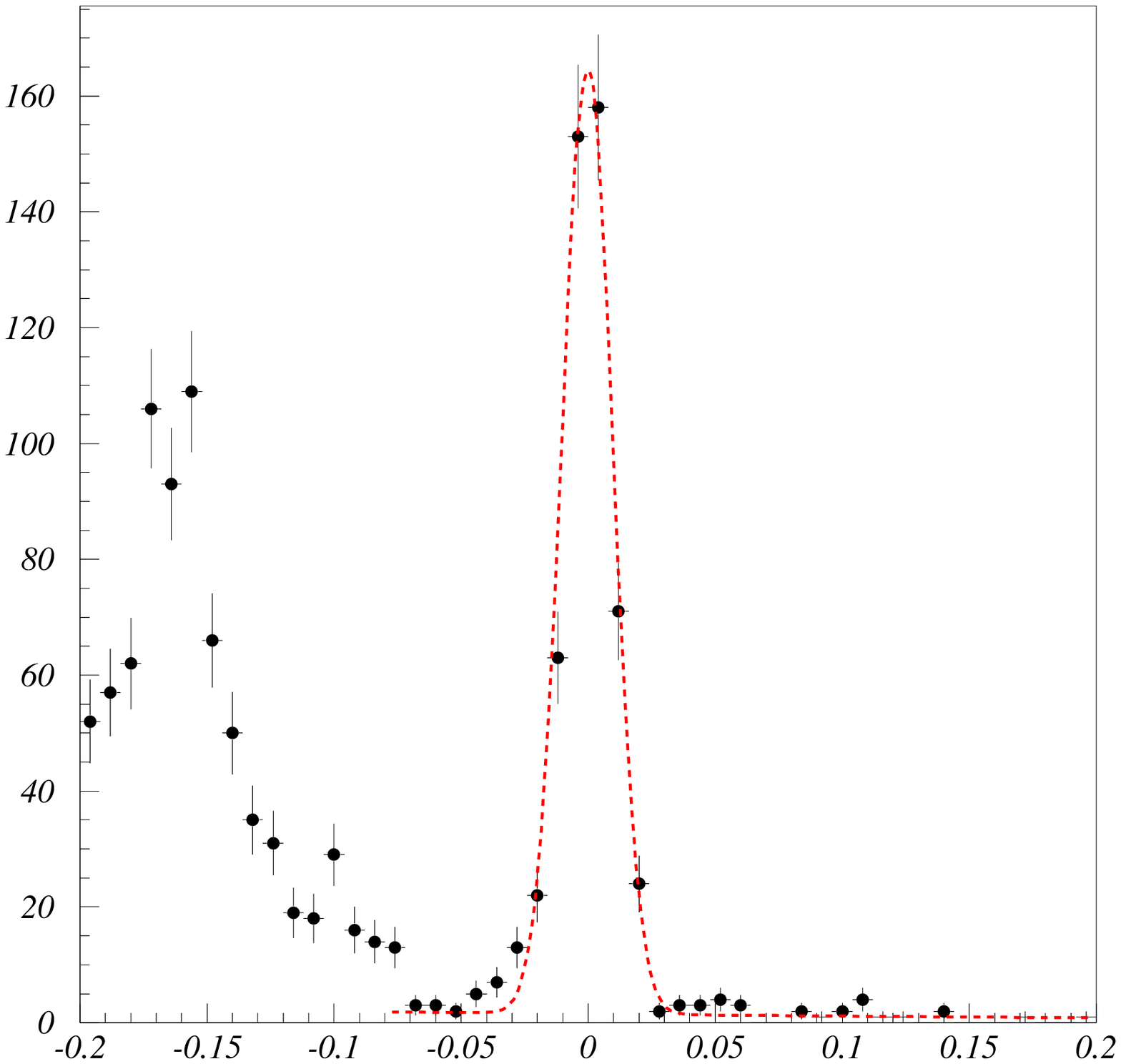}
\end{picture}
\end{center}
\setlength{\unitlength}{1mm}
\begin{center}
\begin{picture}(130,90)
\put(80,0){\large $M_\mathrm{bc}$[GeV/$c^2$]}
\put(10,77){\large\bf $\frac{dN}{dM_\mathrm{bc}\;\cdot\;(0.002\;\mathrm{GeV}/c^2)}$}
\includegraphics[height=8cm,width=12cm]{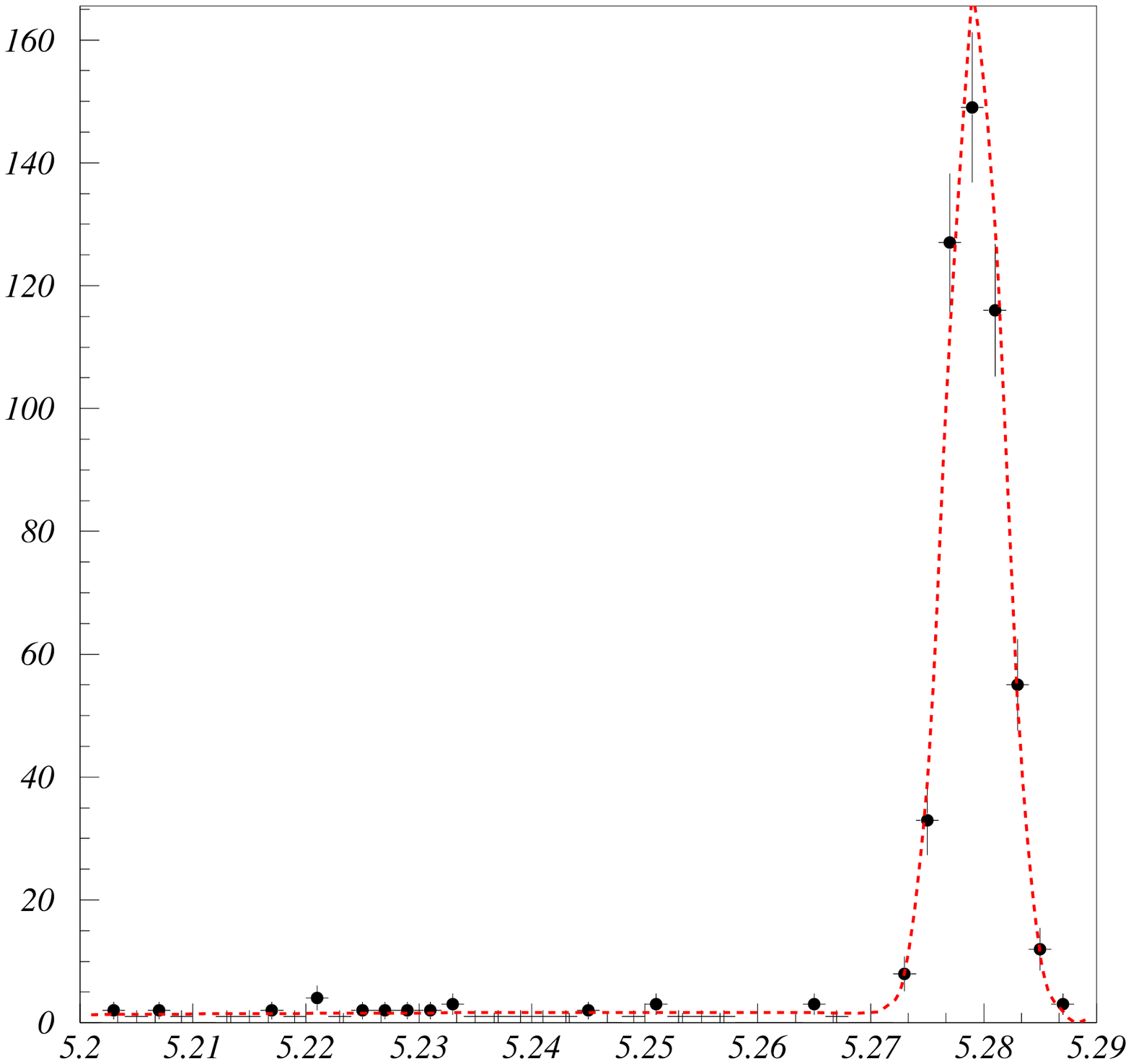}
\end{picture}
\end{center}
\caption{\it Distributions of the $\Delta E$ (top) and $M_\mathrm{bc}$ variables (bottom) for the decay 
$B^{+} \rightarrow D_s^{+} \bar{D^0}$, $D_s^+\to \phi \pi^+$. Points
correspond to data and the dashed (red) line represents results of the fit described in the text. }
\label{FIG_DSD0_1}
\end{figure}

\begin{figure}[t]
\setlength{\unitlength}{1mm}
\begin{center}
\begin{picture}(130,90)
\put(88,0){\large $\Delta E$[GeV]}
\put(10,77){\large\bf $\frac{dN}{d(\Delta E)\;\cdot\;(0.008\;\mathrm{GeV})}$}
\includegraphics[height=8cm,width=12cm]{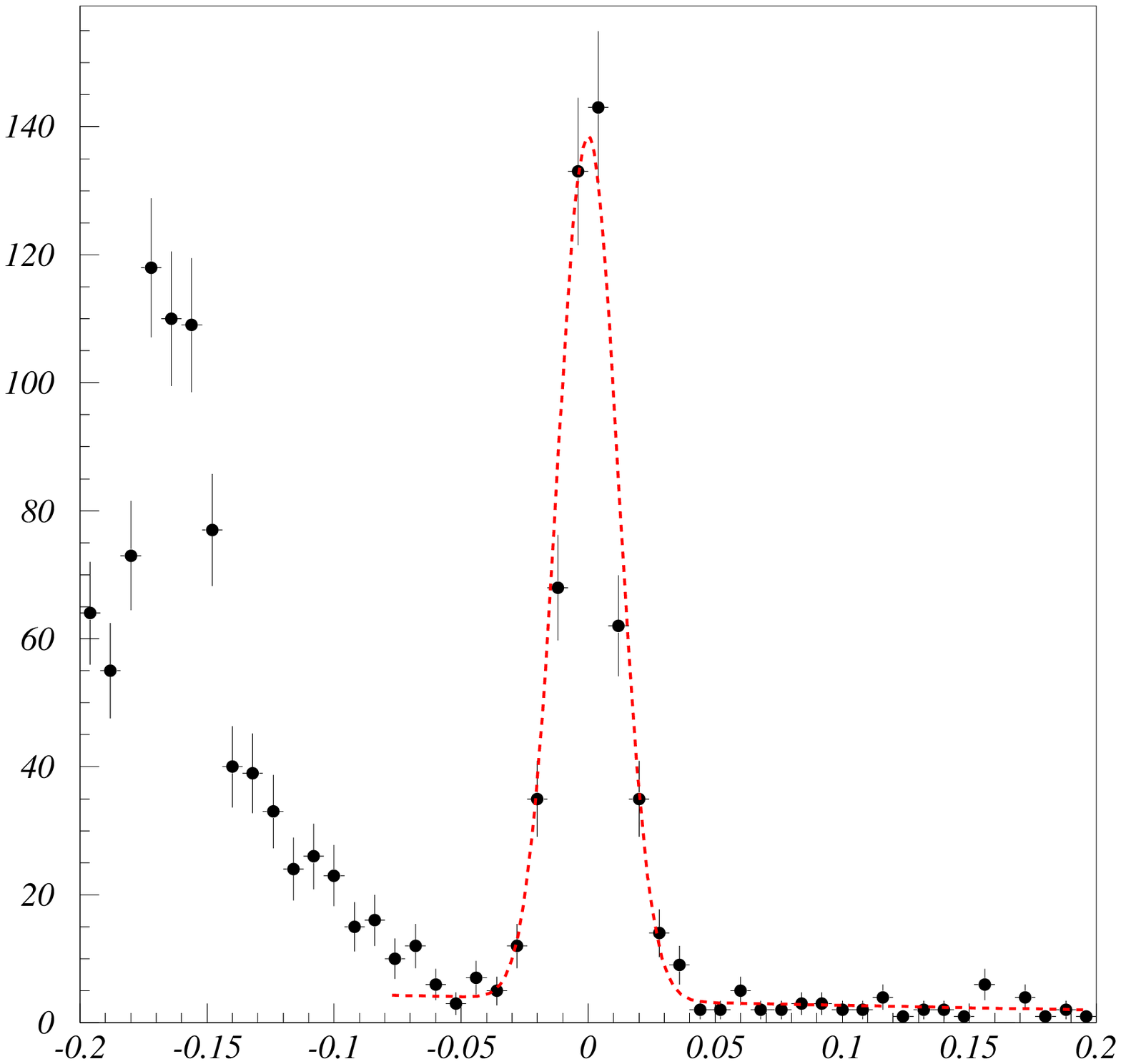}
\end{picture}
\end{center}
\setlength{\unitlength}{1mm}
\begin{center}
\begin{picture}(130,90)
\put(80,0){\large $M_\mathrm{bc}$[GeV/$c^2$]}
\put(10,77){\large\bf $\frac{dN}{dM_\mathrm{bc}\;\cdot\;(0.002\;\mathrm{GeV}/c^2)}$}
\includegraphics[height=8cm,width=12cm]{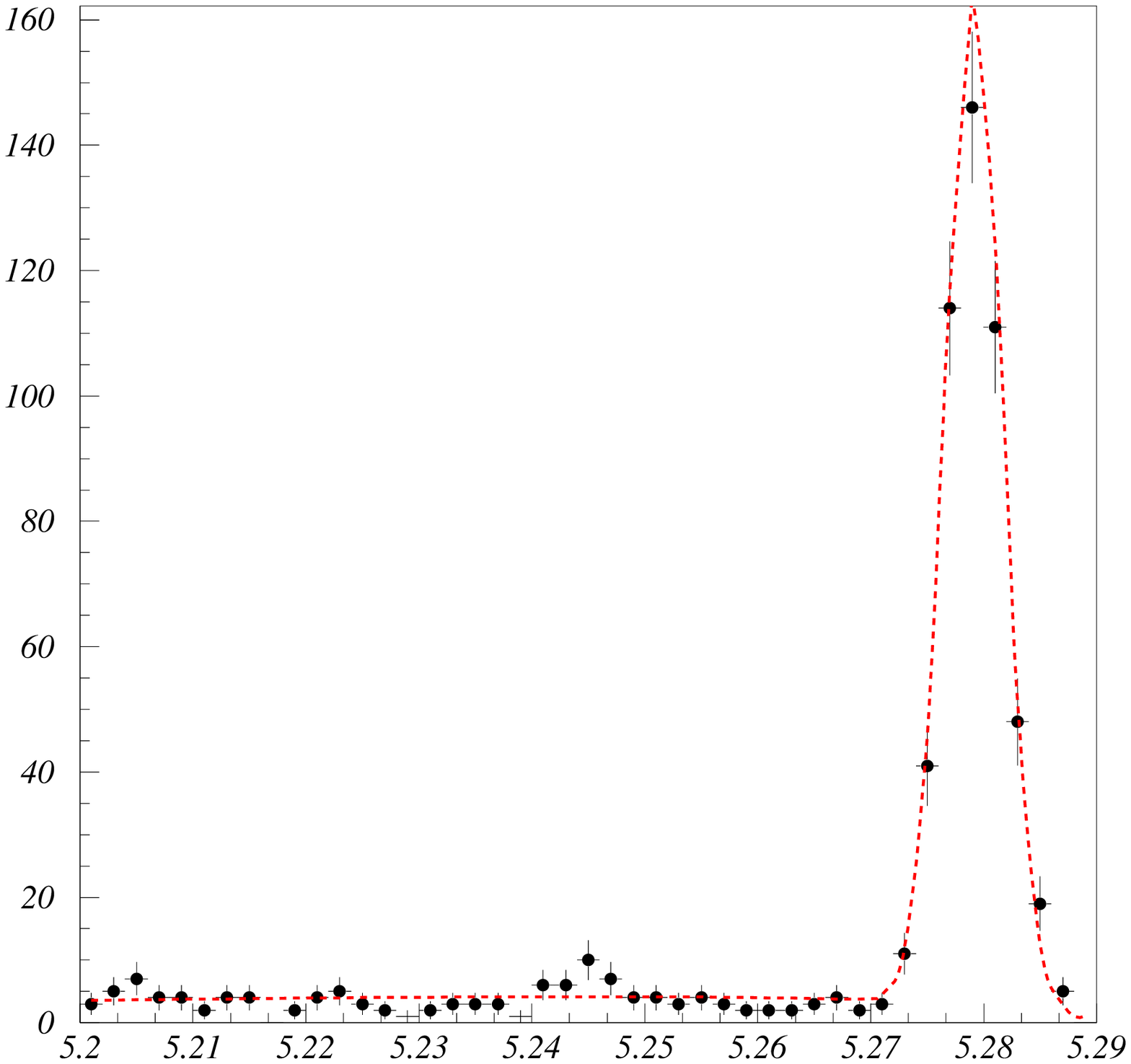}
\end{picture}
\end{center}
\caption{\it Distributions of the $\Delta E$ (top) and $M_\mathrm{bc}$ variables (bottom) for the decay 
$B^{+} \rightarrow D_s^{+}\bar{D^0} $, $D_s^-\to \bar{K^*}(892)^0
K^+$. Points  correspond to data and the dashed (red) line represents
results of the fit described in the text. }
\label{FIG_DSD0_2}
\end{figure}

For events passing all selection criteria mentioned above the $\Delta
E$ distributions (requiring $M_\mathrm{bc} > 5.27$ GeV/c$^2$) and $M_\mathrm{bc}$
distributions (requiring $|\Delta E| < 0.03$ GeV) are examined. The
plots shown in Figs.~\ref{FIG_DSKAPI_1}--\ref{FIG_DSD0_2} exhibit
clear signals from the corresponding $B$ meson decays. In addition, an
excess of events can be seen in the range $\Delta E \in (-0.2,
-0.08)$~GeV, probably due to events with one unreconstructed particle,
such as a $\pi^0$ or $\gamma$, in the final state.


\section{Determination of $B \rightarrow D_s K \pi$ branching fraction}


The branching fractions are calculated using the formula:
\begin{equation}
  {\cal B}(B \rightarrow D_s K\pi) = 
  \frac{N_S/\varepsilon}{N_{B\bar{B}}\cdot {\cal B}_{int}}
  \label{WZOR2}
\end{equation}
where $N_{B\bar{B}}$ is the number of produced $B\bar{B}$ pairs, $N_S$
is the number of reconstructed $B \rightarrow D_s K\pi$ events,
$\varepsilon$ denotes the reconstruction efficiency and ${\cal
B}_{int}$ is the product of branching fractions of intermediate
resonances, present in the respective $B \rightarrow D_s K\pi$
processes (values summarised in Table \ref{TABREZ} are taken from
the Ref.~\cite{PDG}).
\begin{table}[htb]
  \caption{ Branching fractions of intermediate resonances present in 
    $B^+ \rightarrow D_s^{-} K^+\pi^+$ and $B^+ \rightarrow D_s^{+}\bar{D^0}$ decays (from the PDG).}
  \label{TABREZ}
  \begin{tabular}
    {@{\hspace{0.5cm}}l@{\hspace{0.5cm}}|@{\hspace{0.5cm}}c@{\hspace{0.5cm}}}
    \hline \hline
    Decay                                                                              & Branching fraction [\%] \\
    \hline
    $D_s^+ \rightarrow \phi\pi^+, \;\;\phi \rightarrow K^+K^-$                         & 2.16 $\pm$ 0.28         \\
    $D_s^+ \rightarrow \bar{K^*}(892)^0 K^+,\;\;\bar{K^*}(892)^0 \rightarrow K^-\pi^+$ & 2.5 $\pm$ 0.5           \\
    $D^0 \rightarrow K^-\pi^+$                                                         & 3.80 $\pm$ 0.07          \\
    \hline \hline
  \end{tabular}
\end{table}

\subsection{Determination of the number of $B \rightarrow D_s K\pi$ events}

The yields of $B \rightarrow D_s K\pi$ events are determined from an
unbinned extended maximum likelihood fit to $(\Delta E, M_\mathrm{bc},
m_{D_S})$ distributions using the MINUIT~\cite{MINUIT} package. The
likelihood function is given by:
\begin{equation*}
  {\cal L}(\vec \theta)=\prod_{i=1}^{N} f(\vec x_i|\vec \theta) \, ,
\end{equation*}
where $\vec x_i$ is a vector of independently measured values of the
probability density function $f(\vec x|\vec \theta)$. The index $i$
($i$ = 1, 2, $\ldots$, $N$) counts the number of reconstructed events
used in the fit. The vector $\vec \theta=(\theta_1,\ldots, \theta_n)$
contains the parameters that are obtained by maximizing the likelihood
function:
\begin{equation*}
  \hat{\vec \theta} = \max_{\vec \theta}{\cal L}(\vec \theta)
\end{equation*}
The probability density function $f$ contains a Poisson factor to include
the relation between the estimated number of all events ($N_S + N_B$) and the
number of reconstructed events $N$, and is defined as:
\begin{equation*}
  f(\vec x|\vec \theta) = 
  {g(\vec x|\vec \theta)} \cdot 
  \frac{1}{N!}(N_S+N_B)^N e^{-N_S-N_B} \, ,
\end{equation*}
while 
\begin{align}
  g(\vec x|\vec \theta) 
  &= \frac{N_S}{N_S+N_B} \cdot 
     \frac{1}{\sigma_{\Delta E} 
     \sqrt{2\pi}}\exp\biggl[-\frac{\Delta E^2}{2{\sigma_{\Delta E}}^2}\biggr] \cdot 
   \tag{a}\\
  & \qquad\qquad \frac{1}{\sigma_{M_\mathrm{bc}} 
    \sqrt{2\pi}}\exp\biggl[ -\frac{\left(M_\mathrm{bc}-m_{WA}(B)\right)^2}{2{\sigma_{M_\mathrm{bc}}}^2}\biggr] \cdot 
   \tag{b}\\
  & \frac{1}{\sqrt{2\pi}}
    \left(\frac{1}{\sigma_{m_{D_S1}}}\exp\biggl[ -
          \frac{\left(m_{D_S}-m_{WA}(D_s)\right)^2}{2{\sigma_{m_{D_S1}}}^2}\biggr] + 
          \frac{1}{\sigma_{m_{D_S2}}}\exp\biggl[ -
          \frac{\left( m_{D_S}-m_{WA}(D_s)\right)^2}{2{\sigma_{m_{D_S2}}}^2}\biggr]
    \right)
    \tag{c}\\
  & +\frac{N_B}{N_S+N_B} \cdot \frac{1}{\mathcal{N}}\left(w_0+w_1\Delta E+w_2{\Delta E}^2\right) \cdot 
    \tag{d}\\
  & \qquad \frac{M_\mathrm{bc}}{\sqrt{s}/2} \cdot 
    \sqrt{1-\left(\frac{M_\mathrm{bc}}{\sqrt{s}/2}\right)^2}
    \exp\left\lbrace arg\biggl[1-\left(\frac{M_\mathrm{bc}}{\sqrt{s}/2}\right)^2\biggr]\right\rbrace 
    \tag{e}\\
  & \qquad \left( p_0+p_1 m_{D_S}+p_2{m_{D_S}}^2\right)
    \tag{f} \, .
\end{align}
The first three contributions in this formula ((a),(b) and (c))
describe the signal parameterized by four Gaussian functions: one for
$\Delta E$, one for $M_\mathrm{bc}$ and two Gaussians with the same
mean and different widths for the ${m_{D_S}}$ distributions. The last
three terms describe the background: second-order polynomials for
$\Delta E$ and ${m_{D_S}}$ variables ((d) and (f)), and the so-called
ARGUS function~\cite{ARGUS} in (e) for the $M_\mathrm{bc}$ variable.

In this notation $\vec x = (\Delta E,M_\mathrm{bc},m_{D_S})$ and $\vec \theta
= (N_S, N_B, w_0, w_1, w_2, p_0, p_1, p_2, arg, \sigma_{\Delta E}, \\
\sigma_{M_\mathrm{bc}},\sigma_{m_{D_S1}},\sigma_{m_{D_S2}} $). Here $N_S$ and
\begin{table}[b]
  \caption{ Values of the $M_\mathrm{bc}$, $\Delta E$ and $m_{D_S}$ parameters, as extracted from the fits described in the text.}
  \label{TABREZ_MCBDE}
  \begin{tabular}
    {@{\hspace{0.5cm}}l@{\hspace{0.5cm}}|@{\hspace{0.5cm}}c@{\hspace{0.5cm}}c@{\hspace{0.5cm}}c@{\hspace{0.5cm}}}
    \hline \hline
    Decay & $M_{bc}$ [MeV/c$^2$] & $\Delta E$ [MeV] & $m_{D_S}$ [MeV/c$^2$] \\
    \hline
    $B^{+} \rightarrow D_s^{-}(\to\phi\pi^-)K^{+}\pi^{+}$                  & $5279.4\pm 0.2$ & $ -1.3\pm 0.7$ & $1967.8\pm 0.3$ \\
    $B^{+} \rightarrow D_s^{-}(\to K^*(892)^0 K^-)K^{+}\pi^{+}$            & $5279.2\pm 0.2$ & $ -2.1\pm 0.9$ & $1968.7\pm 0.4$ \\
    $B^{+} \rightarrow D_s^+(\to \phi\pi^+)\bar{D^0}$                      & $5279.2\pm 0.1$ & $ -0.3\pm 0.5$ & $1968.2\pm 0.2$ \\
    $B^{+} \rightarrow D_s^+(\to \bar{K^*}(892)^0 K^+)\bar{D^0}$           & $5279.1\pm 0.1$ & $ -0.2\pm 0.6$ & $1967.9\pm 0.2$ \\
    \hline \hline
  \end{tabular}
\end{table}
$N_B$ denote the number of signal events and the number of background
events, respectively, while $w_0, w_1, w_2, p_0, p_1, p_2 $ are the
parameters of the polynomials, $arg$ is a parameter of the ARGUS
function and $\sigma_{\Delta E}$, $\sigma_{M_\mathrm{bc}}$,
$\sigma_{m_{D_S1}}$, $\sigma_{m_{D_S2}}$ are the respective widths of
the Gaussian functions. The factor $\mathcal{N}$ provides a
normalization.

When performing the fit, all $\vec \theta$ parameters are allowed to
float. The mean values of $\Delta E$, $M_\mathrm{bc}$ and $m_{D_S}$ obtained
from the fit are collected in Table~\ref{TABREZ_MCBDE}. The quoted 
$M_\mathrm{bc}$ ($m_{D_S}$) values are mutually consistent and
also in agreement with the world average value $m_{WA}(B)$ ($m_{WA}(D_S)$)~\cite{PDG}.

The determination of the number of reconstructed decays ($N$) proceeds in two steps:
\begin{itemize}
\item 
  The fit to the ($\Delta E$, ${M_\mathrm{bc}}$, $m_{D_S}$)
  distribution is evaluated in the signal region (with requirements
  on all intermediate resonances), resulting in $N_0$ events.
\item 
  For decays containing $K^*(892)^0$ meson, a fit to the sample of
  events from the $K^*(892)^0$ mass sidebands ([0.746,0.796] GeV/$c^2$
  and [0.996,1.046] GeV/$c^2$) is also performed. It yields the number
  of background events, $N_1$, peaking in the signal region and
  therefore contributing to the $N_0$ value. The same procedure is
  applied also for the efficiency calculation. This step is motivated by
  a MC study, where such a background contribution was found. The study
  also revealed that for the $D_s \to \phi\pi$ decay mode this
  background contribution is negligible.
\end{itemize}

Finally, the number of $B \rightarrow D_s K\pi$ events is calculated as:
\begin{equation}
N = N_0 - N_1.
\label{WZORN}
\end{equation}
The values of $N_0$, $N_1$ and $N$ for the analysed decays are listed in Table~\ref{FIGSIG}.

\begin{table}[htb]
  \caption{ Numbers of the signal and background events for analysed decays and efficiencies of their reconstructions. The errors are statistical only.}
  \label{FIGSIG}
  \begin{tabular}
    {@{\hspace{0.5cm}}l@{\hspace{0.5cm}}|@{\hspace{0.5cm}}c@{\hspace{0.5cm}}c@{\hspace{0.5cm}}c@{\hspace{0.5cm}}c@{\hspace{0.5cm}}}
    \hline \hline
    Decay & $N_0$ & $N_1$ & $N$ & $\varepsilon$ [\%] \\
    \hline
    $B^{+} \rightarrow D_s^{-}(\phi\pi)K^{+}\pi^{+}$             & $276.5\;^{+19.3}_{-18.6}$ & -                         & $276.5\;^{+19.3}_{-18.6}$ & $13.89 \pm 0.13$ \\
    $B^{+} \rightarrow D_s^{-}(K^{\star 0} K^-)K^{+}\pi^{+}$     & $299.5\;^{+26.0}_{-25.1}$ & $48.2\;^{+18.6}_{-13.7}$  & $251.3\;^{+32.0}_{-28.6}$ & $9.05 \pm 0.11$  \\
    $B^{+} \rightarrow D_s^{+}(\phi\pi)\bar{D^0}$                & $512.3\;^{+23.4}_{-22.8}$ & -                         & $512.3\;^{+23.4}_{-22.8}$ & $14.09 \pm 0.13$ \\
    $B^{+} \rightarrow D_s^{+}(\bar{K}^{\star 0} K^+) \bar{D^0}$ & $508.0\;^{+24.1}_{-23.4}$ & $47.8\;^{+10.4}_{-7.6}$   & $460.2\;^{+26.2}_{-24.6}$ & $10.03 \pm 0.12$ \\
    \hline \hline
  \end{tabular}
\end{table}

\begin{table}[htb]
  \caption{Values of contributions (in \%) to the overall systematic
           uncertainty in the branching fraction for the $B^+ \rightarrow D_s^{-}
           K^+\pi^+$ decay.}
  \label{FIGSYS}
  \begin{tabular}
    {@{\hspace{0.5cm}}l@{\hspace{0.5cm}}|@{\hspace{0.5cm}}c@{\hspace{0.5cm}}c@{\hspace{0.5cm}}}
    \hline \hline
    Individual                      & Values [\%] for studied decays:                    &                                                          \\
    contribution                    & $B^{+} \rightarrow D_s^{-}(\phi\pi^-)K^{+}\pi^{+}$ & $B^{+} \rightarrow D_s^{-}(K^{\star 0} K^-)K^{+}\pi^{+}$ \\
    \hline
    $\Delta {\cal B}^{\varepsilon}$ & 5.05 & 4.16 \\
    $\Delta {\cal B}^{rng}$         & 1.26 & 2.39 \\
    $\Delta {\cal B}^{\Gamma}$      & 1.57 & 8.04 \\
    $\Delta {\cal B}^{sel}$         & 2.43 & 0.72 \\
    $\Delta {\cal B}^{id}$          & 5    & 5    \\
    $\Delta {\cal B}^{trck}$        & 5    & 5    \\
    $\Delta {\cal B}^{B\bar{B}}$    & 1.3  & 1.3  \\
    \hline
    total                           & 9.34 & 11.83\\
    \hline \hline
  \end{tabular}
\end{table}

\subsection{Determination of the reconstruction efficiency}

The reconstruction efficiency ($\varepsilon$) of each channel in
question is determined using MC samples of $10^5$ $e^+e^- \rightarrow
\Upsilon(4S) \rightarrow B^+B^-$ decays. These samples are generated
in two subsets corresponding to the two inner detector configurations that
were used in Belle experiment (denoted as SVD1 and SVD2). Each subset 
contains $5 \times 10^4$ events. In each event, one of the charged $B$
mesons decays to the appropriate $D_s^{\mp} K^\pm\pi^\pm$ ($D_s^{+}
\bar{D^0}$/$D_s^{-} D^0$) final state. These dedicated MC samples are
subjected to the same analysis as the data, yielding the number of
reconstructed events, $N_{MC_1}$ and $N_{MC_2}$, which corresponds to
the SVD1 and SVD2 samples, respectively. These values divided by the
number of all simulated events, gives partial reconstruction
efficiencies for the studied decays:
\begin{equation*}
  \varepsilon_1 = \frac{N_{MC_1}}{50000}, \qquad \varepsilon_2 = \frac{N_{MC_2}}{50000}
\end{equation*}
The final efficiences for all studied decays are obtained by
calculating the weighted averages of the values defined above:
\begin{equation*}
  \varepsilon = 
  \frac{N_{B \bar{B}1} \cdot \varepsilon_1 + N_{B \bar{B}2} \cdot \varepsilon_2}
       {N_{B \bar{B}1} + N_{B \bar{B}2}}
\end{equation*}
where $N_{B \bar{B}1}$ and $N_{B \bar{B}2}$ are the numbers of $B
\bar{B}$ events collected during the SVD1 and SVD2 running periods,
respectively. Values of efficiencies for studied decays are presented
in Table~\ref{FIGSIG}.

\subsection{Studies of systematic uncertainties}

The following sources of systematic uncertainties are taken into account
(cf. Table \ref{FIGSYS}):
\begin{itemize}

\item 
  $\Delta {\cal B}^{\varepsilon}$ - uncertainty of efficiency
  determination estimated as a statistical error of 
  its measurement, i.e.   
  \begin{equation*}
    \Delta \varepsilon = 
    \sqrt{\left(\frac{N_{B\bar{B}1}}{N_{B \bar{B}1} + N_{B\bar{B}2}}\right)^2
          \cdot (\Delta \varepsilon_1)^2 + 
          \left(\frac{N_{B\bar{B}2}}{N_{B \bar{B}1} + N_{B\bar{B}2}}\right)^2
          \cdot (\Delta \varepsilon_2)^2}
  \end{equation*}
  where $\Delta \varepsilon_1$ and $\Delta \varepsilon_2$ are
  statistical uncertainities on partial efficiencies.\\ Considering the
  control channel and taking into account a mismatch between branching
  fractions calculated for the MC sample (cf. IV.D subsection), an
  additional contribution to the efficiency systematics can be
  assigned. The shift (with an error) between the obtained branching
  fraction value and the generator value is calculated and the corresponding
  systematic error is determined. This factor is summed in quadrature
  with the above $\Delta \varepsilon$ value to evaluate the final
  $\Delta {\cal B}^{\varepsilon}$ contribution.

\item 
  $\Delta {\cal B}^{sel}$ - uncertainty due to the selection
  procedure. To estimate this error, the requirement for the $R_2$
  parameter is varied: $R_2 < 0.40 \rightarrow R_2 < 0.35$.
\item    
  $\Delta {\cal B}^{rng}$ - error from changing the range of the fit
  to the $\Delta E$ variable. For $B^{+} \rightarrow
  D_s^{-}K^{+}\pi^{+}$ decays this range is changed from $(-0.08, 0.2)$
  GeV to $(-0.12, 0.2)$ GeV. For each case the number of signal events
  $N_0^{rng}$ is determined, and the deviation from the $N_0$ value is
  calculated.
\item    
  $\Delta {\cal B}^{\Gamma}$ - uncertainty in the signal
  width. The width of the signal peak, as obtained from the 
  fit to the $\Delta E/M_\mathrm{bc}$ using MC sample, is
  then used as a fixed parameter in the same fit to data. 
  The fixing procedure is applied sequentially: for  $\Delta E$, for
  $M_\mathrm{bc}$ and for both the $\Delta E$ and $M_\mathrm{bc}$ 
  signal peaks. Finally, the maximum deviation from the $N_0$ value is chosen.
\item    
  $\Delta {\cal B}^{id}$ - uncertainty of particle identification 
  in Belle experiment. A standard value of 1\% per charged particle is
  assumed and uncertainties from kaons and pions are combined
  linearly, thus giving an overall contribution of 5~\%.
\item    
  $\Delta {\cal B}^{trck}$ - uncertainty of track reconstruction. 
  A standard value of 1\% per charged track is assumed giving an
  overall contribution of 5~\%.
\item    
  $\Delta {\cal B}^{B\bar{B}}$ - uncertainty of the number of 
  $B\bar{B}$ mesons used as data sample. Value of 1.3\% is assumed 
  in each studied decay.
\end{itemize}
All contributions are assumed to be independent, hence the overall
systematic error is obtained by summing those contributions in
quadrature. The last three sources of systematic uncertainties are
assumed to be common to all decays studied. All information about
systematic uncertainties is collected in Table \ref{FIGSYS}.


\begin{table}[htb]
  \caption{ Branching fraction of studied decays.}
  \label{FIGBR}
  \begin{tabular}
    {@{\hspace{0.5cm}}l@{\hspace{0.5cm}}|@{\hspace{0.5cm}}c@{\hspace{0.5cm}}|c@{\hspace{0.5cm}}c@{\hspace{0.5cm}}c@{\hspace{0.5cm}}c@{\hspace{0.5cm}}|c@{\hspace{0.3cm}}}
    \hline \hline
    studied & ${\cal B}$ & \multicolumn{4}{|c|}{uncertainties} & signif. \\
    \cline {3-6} 
    decays & $\times 10^{-4}$ & stat.(+) & stat.(-) & syst. & ${\cal B}_{int}$ & [$\sigma$] \\
    \hline
    $B^{+} \rightarrow D_s^{-}(\phi\pi)K^{+}\pi^{+}$ & 1.77 & 0.12 & 0.12 & 0.16 & 0.23 & 27.5 \\
    $B^{+} \rightarrow D_s^{-}(K^{\star 0} K^-)K^{+}\pi^{+}$ & 2.15 & 0.27 & 0.24 & 0.25 & 0.43 & 23.1 \\
    $B^{+} \rightarrow D_s^{+}(\phi\pi) \bar{D^0}$ & 85.17 & 3.89 & 3.77 & - & 11.15 & 46.4 \\
    $B^{+} \rightarrow D_s^{+}(K^{\star 0} K^+) \bar{D^0}$ & 92.89 & 5.29 & 4.97 & - & 18.66 & 42.1 \\
    \hline \hline
  \end{tabular}
\end{table}


\subsection{Discussion of the results}

The values of branching fraction for the decays $B^+ \rightarrow
D_s^{-} K^+\pi^+$ and $B^+ \rightarrow D_s^{+}\bar{D^0}$ are collected
in Table~\ref{FIGBR}. Here the '${\cal B}_{int}$' error is due to
uncertainties in the branching fractions for the decays of intermediate
resonances present in the decay in question (Table~\ref{TABREZ}).
The significance of the signal (Table~\ref{FIGBR}) is evaluated
according to the formula $\sqrt{-2 ln({\cal L}_0/{\cal L})}$, where
${\cal L}$ is the likelihood calculated for the nominal fit and ${\cal
L}_0$ is the respective likelihood function for a fit with the number of signal
events fixed to zero.

The final results for the branching fractions for the channels
studied, according to Table~\ref{FIGBR}, are as follows: 

For $D_s \to \phi(\to K^+K^-)\pi$ decay modes:
\begin{equation}
{\cal B}(B^+\to D_s^- K^+\pi^+) = (1.77^{+0.12}_{-0.12} ({\mathrm stat})   \pm 0.16 ({\mathrm syst})   \pm 0.23 ({\mathrm {\cal B}_{int})})\times 10^{-4}~~
\end{equation}
\begin{equation}
{\cal B}(B^+\to D_s^+ \bar{D^0}) = (8.52^{+0.39}_{-0.38} ({\mathrm stat}))\times 10^{-3}~~~~~~~~~~~~~~~~~~~~~~~~~~~~~~~~~~~~~~~~~~
\end{equation}

\vspace*{0.2cm}

and for $D_s \to \bar{K^*}(892)^0(\to K^-\pi^+) K$ decay modes:

\begin{equation}
{\cal B}(B^+\to D_s^- K^+\pi^+) = (2.15^{+0.27}_{-0.24} ({\mathrm stat})   \pm 0.25 ({\mathrm syst})   \pm 0.43 ({\mathrm {\cal B}_{int})})\times 10^{-4}
\end{equation}
\begin{equation}
{\cal B}(B^+\to D_s^+ \bar{D^0}) = (9.29^{+0.53}_{-0.50} ({\mathrm stat}))\times 10^{-3}~~~~~~~~~~~~~~~~~~~~~~~~~~~~~~~~~~~~~~
\end{equation}

\vspace*{0.2cm}

The results presented here are compatible with the values reported by
the BaBar collaboration~\cite{BABAR}. The branching fraction for the
$B^{+} \rightarrow D_s^{+}\bar{D^0}$ decay is consistent with the
world average value given by Particle Data Group (PDG): ${\cal
B}_{PDG}(B^{+} \rightarrow D_s^+\bar{D^0}$) = ($1.09 \pm 0.27 \%$)
\cite{PDG}.

The same study is also performed for a large statistics MC samples
simulating the following processes: $e^+e^-\to B^+ B^-, B^0\bar{B^0}$
and $q\bar{q}$($q=u,d,s,c$). These samples are subjected to the same
analysis procedure as data; the signal from the decay $B^+\to D_s^+
\bar{D^0}$ is fitted as described in subsection IV.A.  The resulting
branching fractions for the decays $B^+\to D_s^+(\rightarrow
\phi\pi^+)\bar{D^0}$ and $B^+\to D_s^+(\rightarrow
K^{*0}K^+)\bar{D^0}$ are $(9.28^{+0.23}_{-0.22}) \times 10^{-3}$ and
$(8.92^{+0.23}_{-0.22}) \times 10^{-3}$, respectively. They are
compatible with the values assumed in the MC generator: ${\cal
B}(B^+\to D_s^+ \overline{D^0}) = 9.06 \times 10^{-3}$ (Differences
within statistical errors of the extracted MC values are used as
conservative estimates of systematic uncertainties for the
reconstruction efficiencies, as described above.)

\subsection{Studies of two-body subsystems $D_s K$, $D_s\pi$ and $K\pi$}
 
The study of two-body subsystems in the $D_sK\pi$ final states is
driven by two facts. The first one is the lack of resonances in the
$D^+\pi^-$ invariant mass above the value of 2.55 GeV/c$^2$ observed
by Belle~\cite{BDPIPI}, where the respective three-body decay is
$B^+\to D^+\pi^-\pi^-$.  Second, the $D_s^-\pi^+$ and $K^+\pi^+$ pairs
cannot form 'standard' $q\bar{q}$ resonances, so any signal observed
in these subsystems would indicate the presence of some exotic states
such as hybrid mesons, tetraquarks, etc.

\begin{figure}[ht]
\setlength{\unitlength}{1mm}
\begin{center}
\begin{picture}(130,90)
\put(53,0){\large $M_{D_sK}^2$ vs. $M_{D_s\pi}^2$[GeV$^2$/$c^4$]}
\includegraphics[height=10cm,width=12cm]{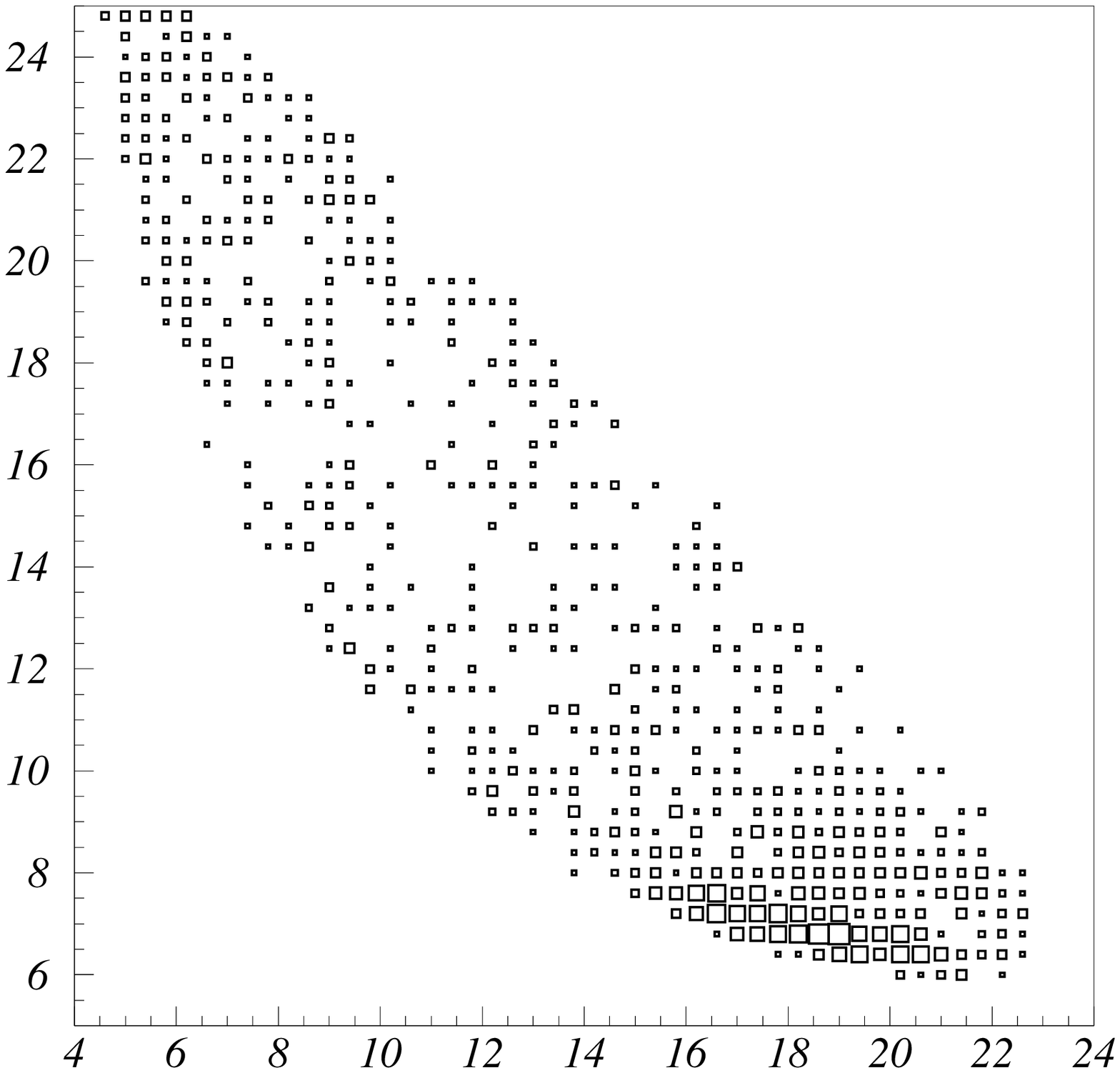}
\end{picture}
\end{center}
\setlength{\unitlength}{1mm}
\begin{center}
\begin{picture}(130,100)
\put(53,0){\large $M_{D_sK}^2$ vs. $M_{D_s\pi}^2$[GeV$^2$/$c^4$]}
\includegraphics[height=10cm,width=12cm]{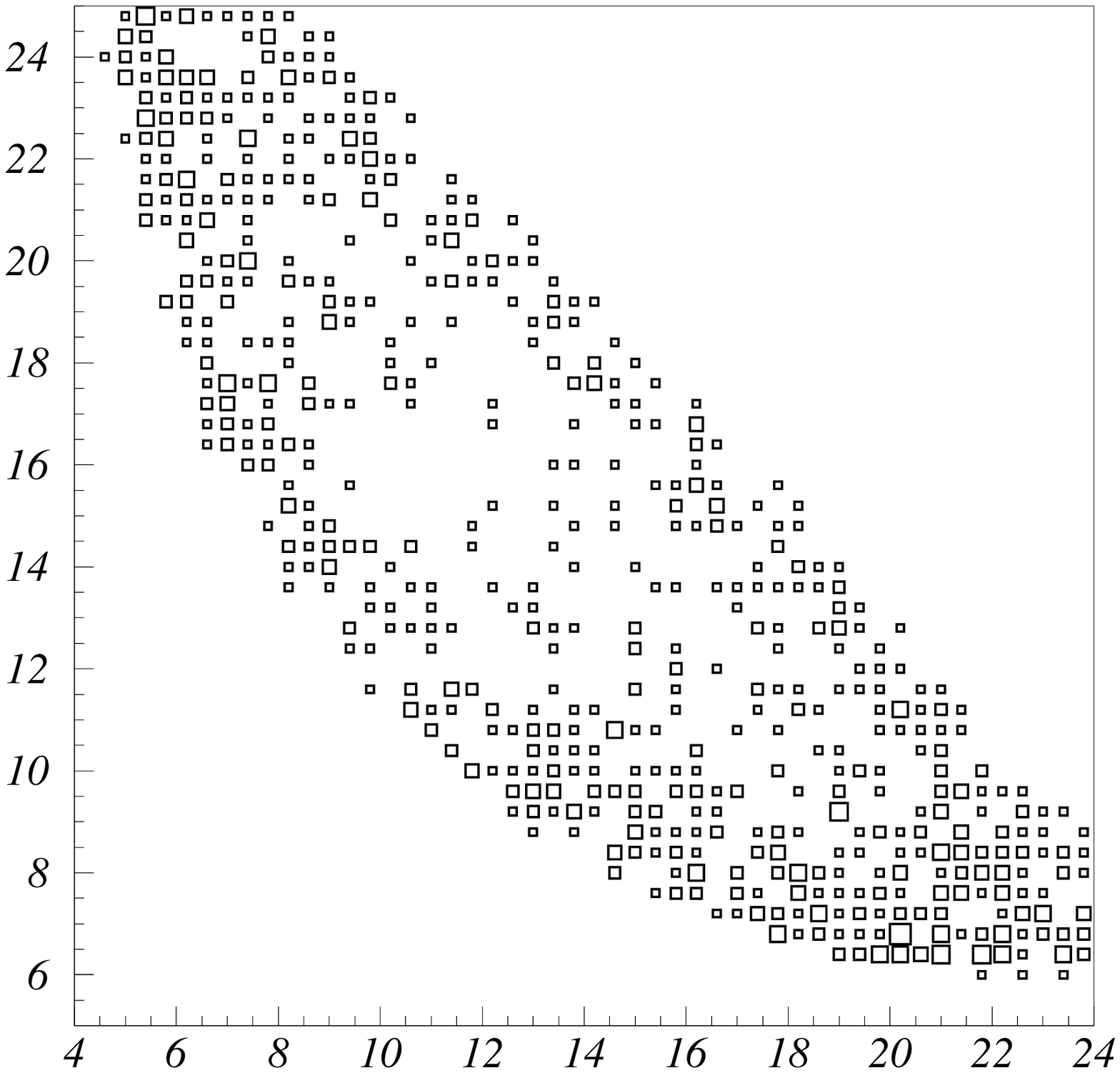}
\end{picture}
\end{center}
\caption{\it The Dalitz plots of $M_{D_sK}^2$ vs $M_{D_s\pi}^2$ for the signal box (upper plot) and $\Delta E$ sidebands (lower plot). }
\label{FIG_DALITZ1}
\end{figure}

\begin{figure}[ht]
\setlength{\unitlength}{1mm}
\begin{center}
\begin{picture}(130,80)
\put(57,0){\large $M_{D_s\pi}^2$ vs. $M_{K\pi}^2$[GeV$^2$/$c^4$]}
\includegraphics[height=10cm,width=12cm]{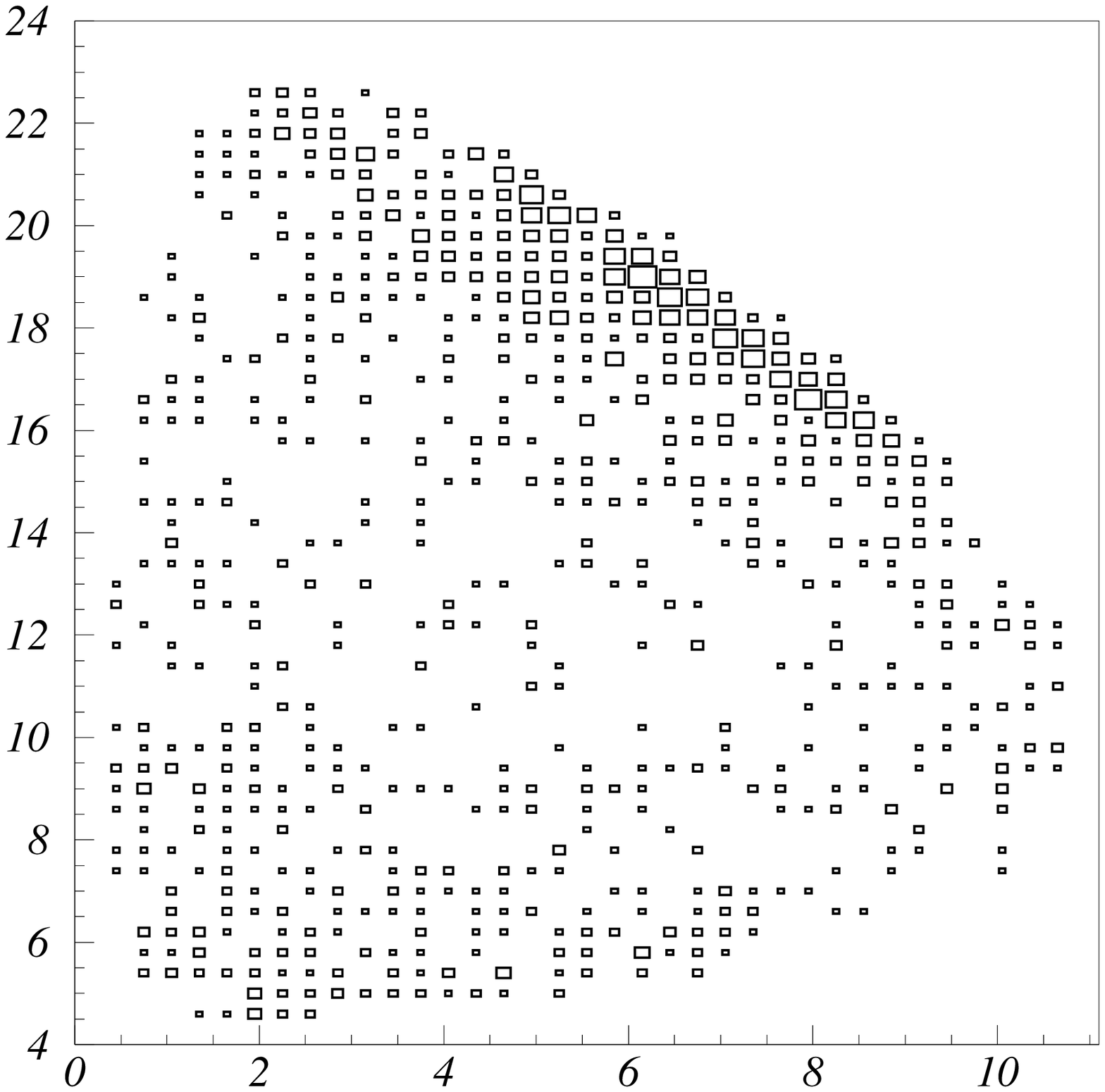}
\end{picture}
\end{center}
\setlength{\unitlength}{1mm}
\begin{center}
\begin{picture}(130,100)
\put(55,0){\large $M_{D_s\pi}^2$ vs. $M_{K\pi}^2$[GeV$^2$/$c^4$]}
\includegraphics[height=10cm,width=12cm]{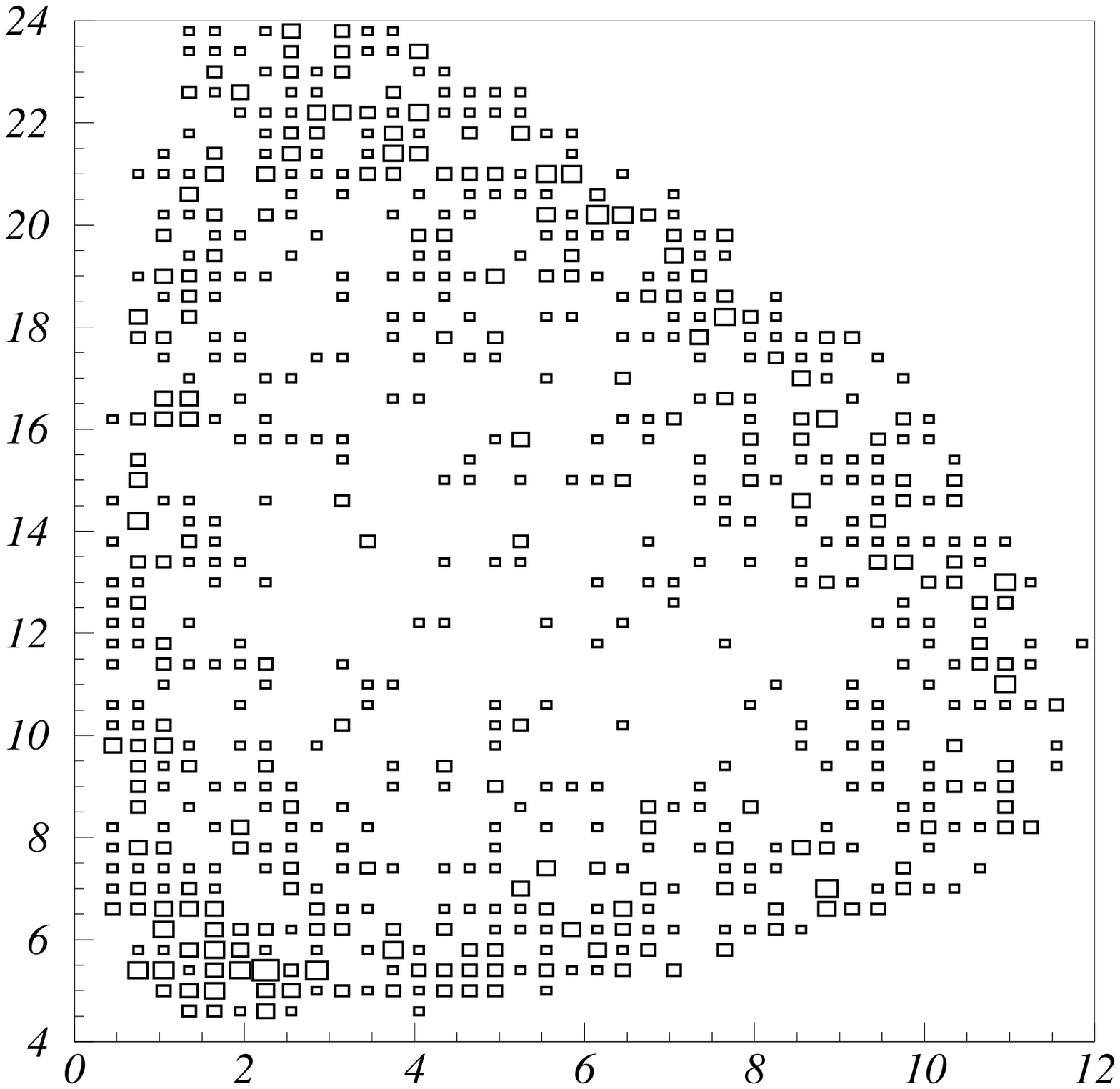}
\end{picture}
\end{center}
\caption{\it The Dalitz plots of $M_{D_s\pi}^2$ vs $M_{K\pi}^2$ for the signal box (upper plot) and $\Delta E$ sidebands (lower plot). }
\label{FIG_DALITZ2}
\end{figure}

\begin{figure}[ht]
\setlength{\unitlength}{1mm}
\begin{center}
\begin{picture}(130,80)
\put(54,0){\large $M_{D_sK}^2$ vs. $M_{K\pi}^2$[GeV$^2$/$c^4$]}
\includegraphics[height=10cm,width=12cm]{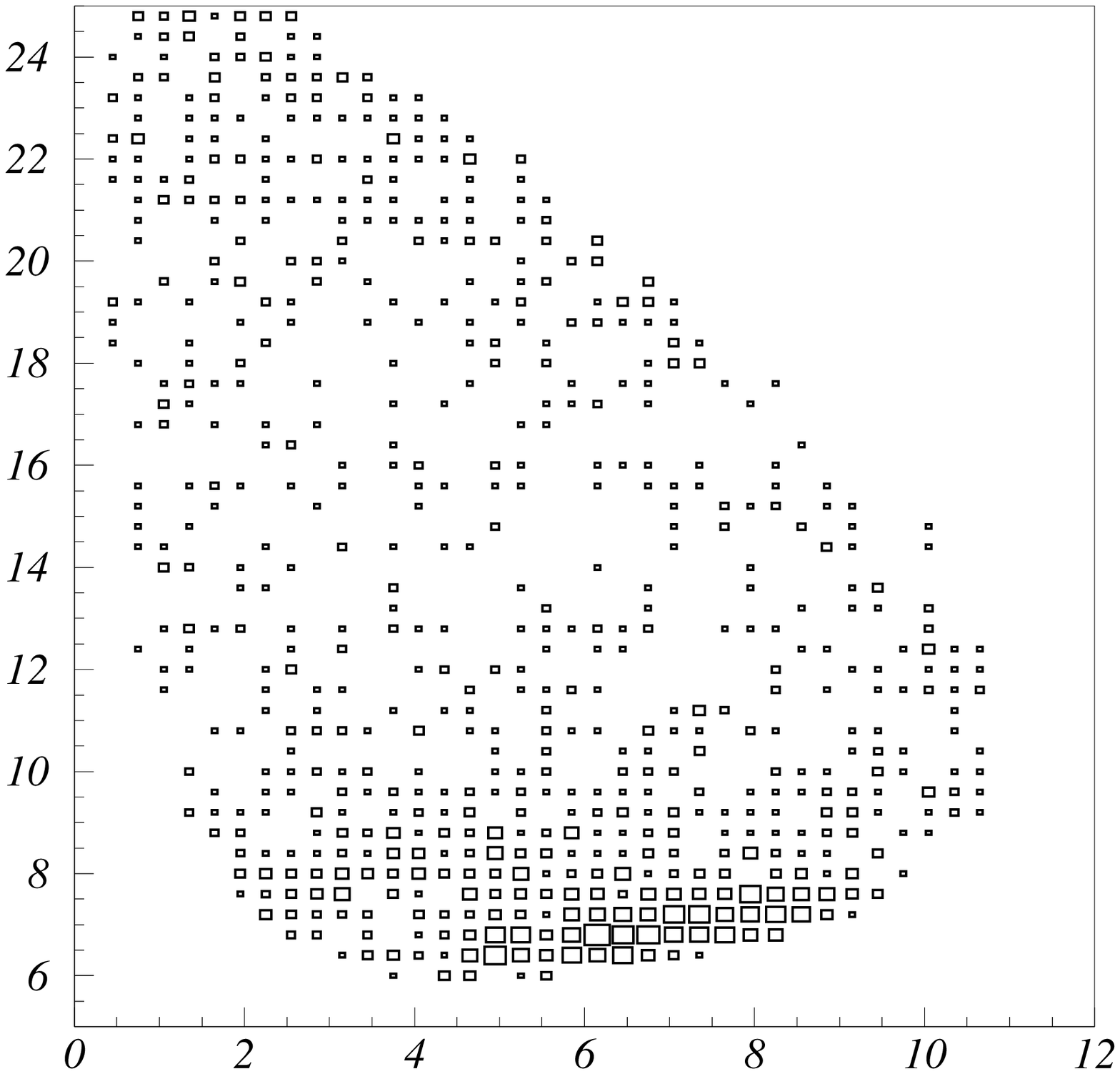}
\end{picture}
\end{center}
\setlength{\unitlength}{1mm}
\begin{center}
\begin{picture}(130,100)
\put(54,0){\large $M_{D_sK}^2$ vs. $M_{K\pi}^2$[GeV$^2$/$c^4$]}
\includegraphics[height=10cm,width=12cm]{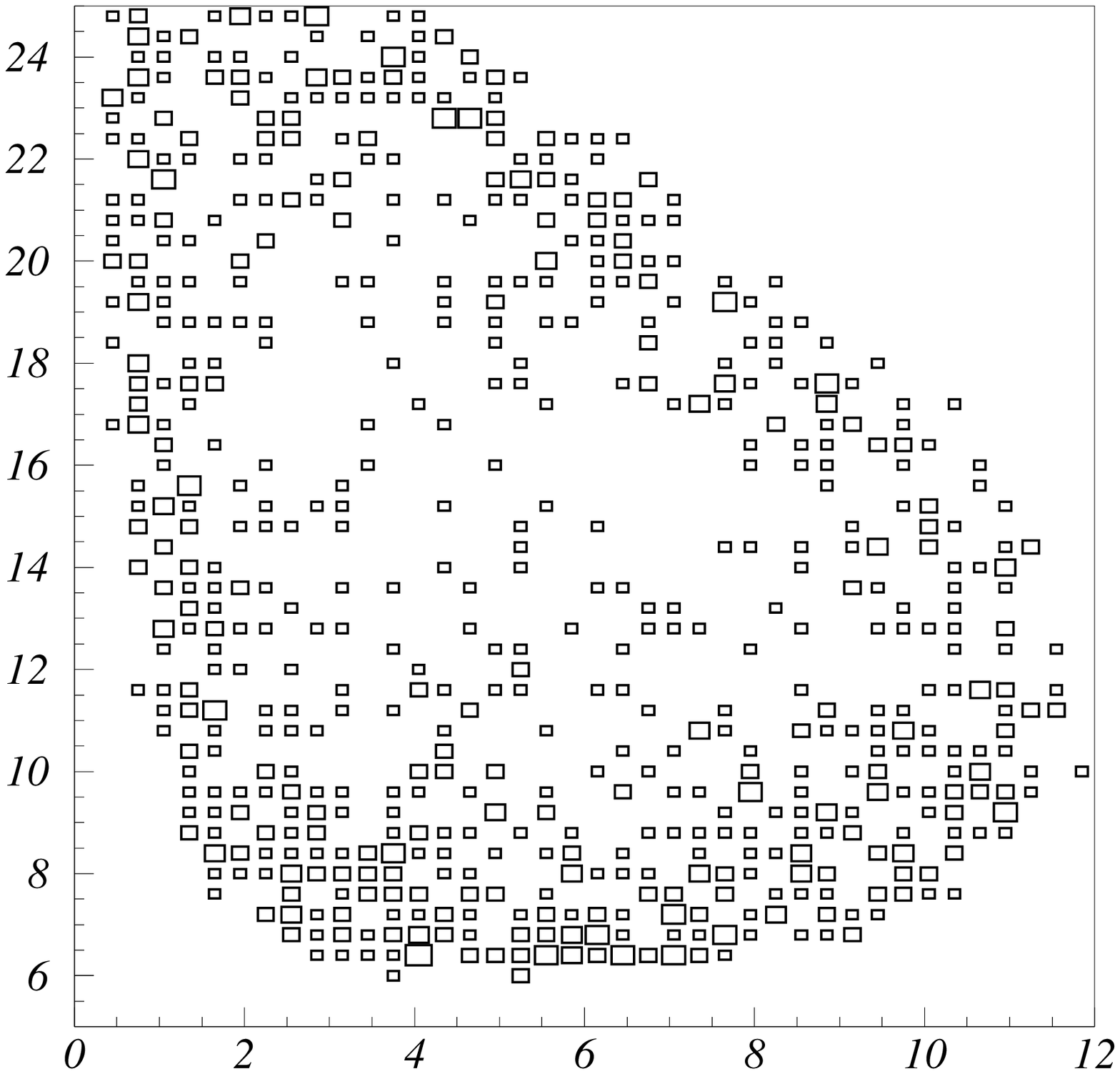}
\end{picture}
\end{center}
\caption{\it The Dalitz plots of $M_{D_sK}^2$ vs $M_{K\pi}^2$ for the signal box (upper plot) and $\Delta E$ sidebands (lower plot). }
\label{FIG_DALITZ3}
\end{figure}

Below we concentrate on the $D_s^-K^+\pi^+$ final
state. Figures~\ref{FIG_DALITZ1}--\ref{FIG_DALITZ3} show the Dalitz
plots of all possible combinations of invariant-mass squared:
$M^2(D_sK)$, $M^2(D_s\pi)$ and $M^2(K\pi)$. All distributions are
shown both for the signal region and the $\Delta E$ sidebands. The
structure visible in the invariant mass of the $D_s\pi$ subsystem in
the mass region around 3~GeV/c$^2$ corresponds to
the decays $B^+\to (c\bar{c}) K^+$. Preliminary studies of these
Dalitz plots do not confirm contributions of any exotic states in the
two-body subsystems. However, further studies might reveal some
enhancements such as that observed by BaBar in the 
$m(D_s^- K^+)$ spectrum \cite{BABAR}. Detailed analyses of the two-body
subsystems are therefore still in progress.


\section{Acknowledgments}

We thank the KEKB group for the excellent operation of the
accelerator, the KEK cryogenics group for the efficient
operation of the solenoid, and the KEK computer group and
the National Institute of Informatics for valuable computing
and Super-SINET network support. We acknowledge support from
the Ministry of Education, Culture, Sports, Science, and
Technology of Japan and the Japan Society for the Promotion
of Science; the Australian Research Council and the
Australian Department of Education, Science and Training;
the National Science Foundation of China and the Knowledge
Innovation Program of the Chinese Academy of Sciences under
contract No.~10575109 and IHEP-U-503; the Department of
Science and Technology of India; 
the BK21 program of the Ministry of Education of Korea, 
the CHEP SRC program and Basic Research program 
(grant No.~R01-2005-000-10089-0) of the Korea Science and
Engineering Foundation, and the Pure Basic Research Group 
program of the Korea Research Foundation; 
the Polish State Committee for Scientific Research; 
the Ministry of Education and Science of the Russian
Federation and the Russian Federal Agency for Atomic Energy;
the Slovenian Research Agency;  the Swiss
National Science Foundation; the National Science Council
and the Ministry of Education of Taiwan; and the U.S.\
Department of Energy.



\begin{thebibliography}{99}
\bibitem{FOOT} 
  Throughout this paper, the inclusion of the charge conjugate mode
  decay is implied unless otherwise stated.
\end{thebibliography}
\end{document}